\documentclass[review]{elsarticle}

\usepackage{lineno,hyperref}
\modulolinenumbers[5]
\usepackage{mhchem}
\usepackage{siunitx} 
\usepackage{amsmath}
\usepackage{natbib}
\usepackage{inputenc}
\usepackage{xcolor}
\usepackage{amsmath}
\usepackage{amsfonts}
\usepackage{mhchem}
\usepackage{amsmath}
\usepackage{amssymb}
\usepackage{latexsym}

\begin{document}


\title{\textcolor{black}{{Tuning Paradigm of External Stimuli Driven Electronic, Optical and Magnetic Properties in Hybrid Perovskites and Metal Organic Complexes}}}

\author{Hrishit Banerjee$\dagger$}
\address{Department of Chemistry, University of Cambridge, Lensfield Road, CB2 1EW, Cambridge, Cambridgeshire, UK.}

\author{Jagjit Kaur$\dagger$}
\address{Materials Theory for Energy Scavenging (MATES) Lab, Department of Physics, Harish-Chandra Research Institute(HRI), A CI of Homi Bhabha National Institute (HBNI), Chhatnag Road, Jhunsi, Prayagraj 211019, India}
\author{M. K. Nazeeruddin\text{*}}
\address{Group for Molecular Engineering of Functional Materials, Institute of Chemical Sciences and Engineering, École Polytechnique Fédérale de Lausanne, Valais Wallis, Sion, Switzerland.}
\ead{mdkhaja.nazeeruddin@epfl.ch}

\author{Sudip Chakraborty\text{*}}
\address{Materials Theory for Energy Scavenging (MATES) Lab, Department of Physics, Harish-Chandra Research Institute(HRI), A CI of Homi Bhabha National Institute (HBNI), Chhatnag Road, Jhunsi, Prayagraj 211019, India\corref{mycorrespondingauthor}}
\cortext[mycorrespondingauthor]{Corresponding author}
\ead{sudipchakraborty@hri.res.in, sudiphys@gmail.com}

\begin{abstract}
We have witnessed a wide range of theoretical as well as experimental investigations to envisage external stimuli induced changes in electronic, optical and magnetic properties in the metal organic complexes, while hybrid perovskites have recently joined this exciting league of explorations. The flexible organic linkers in such complexes are ideal for triggering not only spin transitions  but also a plethora of various different responses under the influence of various external stimuli like pressure, temperature and light. A diverse range of applications particularly in the field of optoelectronics, spintronics and energy scavenging have been manifested. The hysteresis associated with the light induced transitions and spin-crossover governed by pressure and temperature, are promising phenomena for the design principles behind memory devices and optical switches. The pressure induced optical properties tuning or piezochromism has also emerged as one of the prominent areas in the field of hybrid perovskite materials family. It is thus imperative to have a clear understanding of how the tuning in electronic, optical and magnetic properties occur under various stimuli, and selectivity of the stimulus could be influential behind the maximum efficiency in the field of energy and optoelectronic research, and in what future directions this field could be driven from the perspective of futuristic material properties. This review though primarily focuses on the theoretical aspects of understanding the different mechanisms of the phenomena, does provide a unique overview of the experimental literature too, accompanied by the theoretical understanding such that relevant device applications can be considered through a future roadmap of tuning paradigm of external stimuli. It also provides an insight as to how energy and memory storage may be combined by using the principles of spin transition in metal organic complexes.


\end{abstract}

\begin{keyword}
Metal-organic complexes, Hybrid Perovskites, Pressure induced transition, Temperature induced transition, Light Induced Transition, Piezochromism, Spin-crossover
\end{keyword}
\maketitle


\section{Introduction}
Metal organic complexes, \textcolor{black}{which encompasses a very wide spectrum of metal organic molecular systems, metal organic polymers, metal organic frameworks as well as hybrid perovskites} have become centre of attraction in the materials science research due to their exciting properties and possible device applications. \cite{metal-org1, metal-org2, stuart, cote}. The synergistic effect between the flexible and easily synthesised organic part and the mechanically stable inorganic part enables these materials to be promising candidates for optoelectronic  devices with higher carrier mobility as compared to the contemporaries \cite{metal-org1}. With a wide range of possibilities of metal ions and corresponding organic linkers, these complex materials have evolved as a family with diversified structural and chemical properties and therefore are adequately appropriate for materials design towards specific applications. The intrinsic porosity in such materials keep these complexes ahead as far as the applications are concerned in catalysis, chemical sensing, energy storage and biomedical research \cite{sensors, catalysis, biomedical, morris, kreno}. There has been recent progress regarding a new class of materials called hybrid perovskites with less porosity and they are showing promising applications in the diverse area of modern materials science, specially in the area of energy applications like solar cells, light emitting diodes, optical switches and memory devices. The crystal structure of such hybrid perovskites are based on the general formula ABX$_3$, where A is the organic cation, B is the inorganic component and X is the halogen. These materials have emerged as one of the prominent materials family for energy scavenging due to their high absorption coefficient ($\sim$ 10$^{-17}$ cm$^{-2}$) \cite{absorp}, reasonable range of carrier mobilities ranging from $10^{-6}$cm$^2$V$^{-1}$s$^{-1}$ to $10$cm$^2$V$^{-1}$s$^{-1}$ \cite{mobility1, mobility2}, and low exciton binding energy of 2-50 meV \cite{excitonbe} with longer exciton diffusion length of up to 43-87 nm \cite{excitondl} as compared to the organic materials. These materials can also be useful for Light Emitting Diode (LED) applications because of their tunable band gap. The development of the predecessor of such perovskites are the ABO$_3$ oxides, which have also been used in a broad spectrum of  applications as multiferroics, high temperature superconductors, and giant magneto-resistance devices. However, outstanding challenges remain associated with elucidating the fundamental reasons behind all such properties and how they can be exploited to develop higher efficiency devices, guided by computational as well as experimental studies.
\begin{figure}[h!]
	\begin{center}
		\includegraphics[width=\columnwidth]{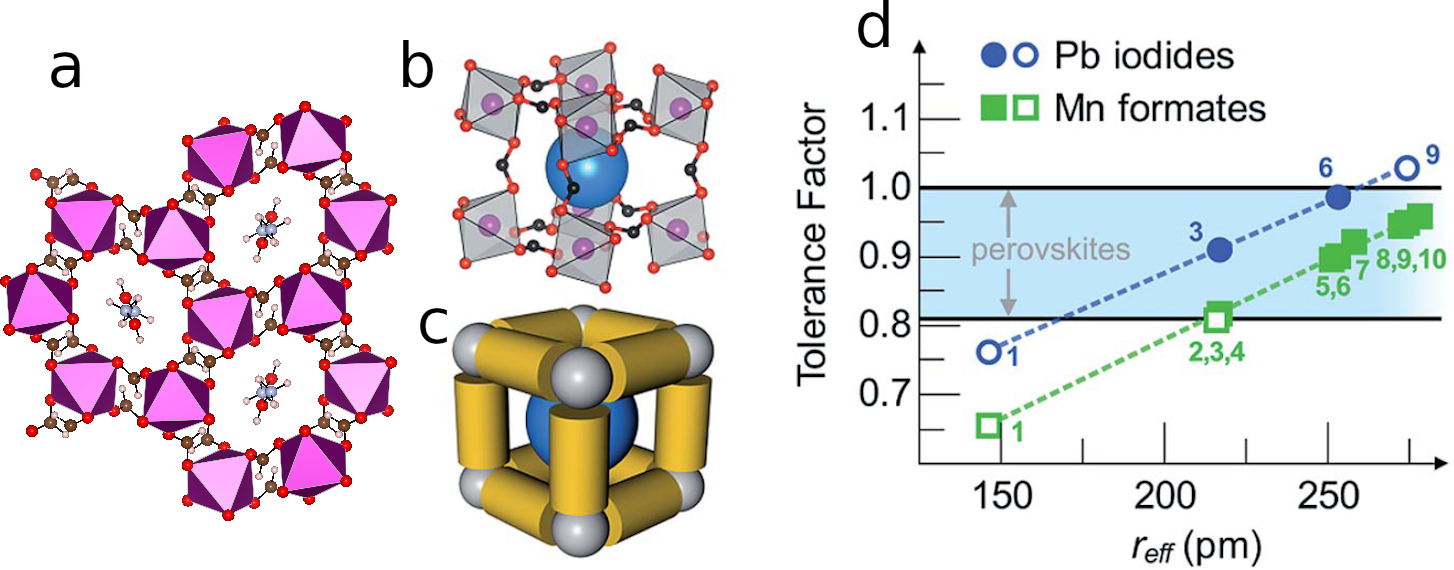}
		\caption{Structure and octahedral connectivity and tolerance factor of [NH$_3$OHMn(HCOO)$_3$, and other Mn based formates in case of tolerance factor demonstrating the amine cation sitting in the perovskite ReO$_3$ type cavity. Panel (a) shows the amine cations at A site sitting in the spherical cavity, (b) shows the octehadral connectivity through the formate bridges, (c) shows a cartoon figure corresponding to the perovskite nature of the material and (d) shows the variation in tolerance factor with $r_{eff}$ for Pb iodides and Mn formates.  Figure partially adopted from Cheetham et al.. \cite{Cheetham}}
	\end{center}
\end{figure}

One of the two important subsets of such hybrid perovskites has [AmH]MX$_3$ composition, where AmH$^+$ is the protonated amine part; M is either Sn$^{2+}$ or Pb$^{2+}$ and X is the halogen part (Cl$^-$, Br$^-$, or I$^-$). The other set of hybrid perovskites are transition metal formates [AmH]M(HCOO)$_3$ (M = Mn, Cu, Ni, Fe, Co), that have recently been synthesised and have shown surprising ferroic properties \cite{hybrid-formate, Clune2020, Sanchez-Andujar2010, Chitnis2018}. The formates are acting like the linkers that connect the MO$_6$ octahedra with the protonated amine molecules being found at the hollow spaces formed by the octahedra. These hollow spaces can act as the pseudo-cubic cavity as shown in Fig. 1. These types of amine based hybrid perovskite systems show multiferroic properties on cooling when accompanied by an order-disorder phase transition along with the disorderedness at ambient temperatures.

Recently,  order-disorder phase transitions along with surprising electronic properties have been found  in these multiferroic materials by Cheetham and co-workers\cite{Cheetham}. There exists a strong correlation between the ionic tolerance factor and the corresponding crystal structure as demonstrated in Fig. 1, that leads to the tuning of electronic properties. This hypothesis has been proven to be correct for several recently synthesised hybrid perovskites \cite{Cheetham}. They have defined exclusively a set of effective ionic radii for the organic and inorganic ions to estimate the tolerance factors for organic-inorganic hybrid perovskites of lead iodides and manganese formates. This has certainly addressed the issue and resolved to a great extent the complex correlation between bonding nature and the crystal engineering of not only hybrid perovskites family, but also other systems like cyanides and azides. They have also put an effort to envisage the agreement between calculated tolerance factors and experimental structural observations thus planting their foot well for future efforts to identify promising combinations consist of amine, metal and anionic parts. But, it is noteworthy to mention here that the tolerance factors for hybrid materials that might have significant effect on the design of materials with specific properties, are not completely conclusive yet. This also might be investigated by applying external strain to the individual perovskite systems and to see how the corresponding electronic, magnetic and optical properties are tuned with the applied hydrostatic pressure exerted on them. The effect of hydrostatic pressure is particularly noteworthy here since pressure is one of the driving mechanisms behind spin crossover \cite{letard, Letard2004, cobo, brooker, halder}. It may also be said that the effect of structure and tolerance factor also gives a nod to the temperature dependent spin crossover studies since heating or cooling a material causes structural changes which effectively leads to spin transitions in materials which can show such transitions. 

There is also an ongoing exchange between theory and corresponding experiments in this exciting new field of metal organic complexes to predict their different inherent properties as a function of external parameters like temperature, pressure, chemical effects or light irradiation. The successful synthesis of formates based perovskites containing amine and metal cations have been performed by Jain and co-workers  and it has been found that most of those systems have shown multiferrocity with order-disorder phase transition at low temperature accompanied by hysteresis and surprising dielectric anomalies \cite{Jain1, Jain2}. A deeper understanding of all such intrinsic properties is therefore needed, which requires a substantial scientific effort to explore extensively and in this context the atomistic insight of these materials is extremely relevant at this point of time.  This paves the way for successful application of electronic structure calculations based on Density Functional Theory (DFT) which has laid much of the foundation of materials modeling by complimenting the experimental investigations to be carried out in the laboratory. Ferroelectricity and especially multiferroicity in these materials has been extensively studied by Stroppa and coworkers mostly from a DFT based first principles perspective and at times combined with experimental studies \cite{stroppa1,stroppa2, stroppa4, stroppa5, Jain2016, sante, Tian2014, aguirre, kamminga, ptak, ghosh, mazzuca} The applicability of such electronic structure calculations is practicable as it has the potential to unveil undiscovered phenomena and properties in systems based on  crystal structure of materials that are available in the process of experimental synthesis. By performing electronic structure calculations, we always include the  structural and chemical aspects of the system in full rigour without loss of generality. 

Spin crossover is one such novel phenomena occurring in above mentioned hybrid frameworks with possibly diverse consequent applications in memory storage, optical switches, and display devices \cite{letard, Letard2004, cobo, brooker, halder, Banerjee2014, Banerjee2016, Banerjee2017}. Banerjee et al have recently demonstrated the cooperative spin-state transitions in Fe and Mn based hybrid perovskites, dimethylammonium iron formate, [CH$_3$NH$_2$CH$_3$][Fe(HCOO)$_3$], and hydroxylammonium iron formate, [NH$_3$OH][Fe(HCOO)$_3$], and dimethylammonium manganese formate [CH$_3$NH$_2$CH$_3$][Mn(HCOO)$_3$] under external stimuli from the first principles electronic structure calculations \cite{Banerjee2016, banerjee2021}. In addition to a large hysteresis, both the materials undergo spin-state transition (from high-spin [S = 2] to low-spin [S = 0]) at Fe (II) site on application of external hydrostatic pressure. This spin state transition can be strongly correlated to the variation of electronic, magnetic, and optical properties. This work was a continuation of a previously performed first principles investigation of temperature driven spin crossover \cite{Banerjee2014} in metal-organic complexes. These investigations certainly open up novel functionalities in such hybrid perovskites family and leading to their possible applications in modern day electronic devices.

The above mentioned searches on spin-crossover driven mainly by external temperature and pressure, are not only confined to one particular scientific community but it has propagated among various scientific communities with a multidimensional interdisciplinary nature and has engaged theorists, experimentalists and technologists involved in device design alike. It is also important to mention here that high throughput calculations may also be of paramount importance in the search for new materials for spin crossover particularly for hybrid perovskites. High throughput calculations have already been used for designing lead free emergent hybrid perovskites for solar cells \cite{Chakraborty2017}. Therefore, it seems imperative to have the relevant previous and ongoing experimental and theoretical results regarding tunabilityy of hybrid perovskites and metal organic complexes in a single coherent nutshell because of the sheer multitude of properties demonstrated by these materials and numerous applications that are linked to those tunable properties. In this review, we have tried to fill the void by having all the relevant information in one place, in a concise comprehensive manner, to have a better understanding of the current status and progress of external stimuli driven (pressure, temperature, and light) spin-crossover in metal organic complexes and hybrid perovskites while trying to provide new ideas for tunability based applications in metal organic complexes and hybrid perovskites alike. Here, we will outline an attempt to introduce a systematic overview, which will be of great current interest and imminent need in the hybrid perovskites based materials, that may show spin crossover phenomena under external pressure and temperature. After providing a comprehensive summary of the experimental investigations in the huge and complicated field of metal organic complexes, we will discuss about the pressure and temperature effect as seen by experimental studies in the upcoming sections. With this background, we will be focussing our discussion on the theoretical studies that have been carried out on hybrid perovskites in particular and spin crossover phenomena in general under the effect of pressure and temperature. The mechanical stability and the corresponding thermal properties of such metal organic complexes are also worth addressing. The microscopic understanding of the spin crossover phenomena and how it takes place in hybrid perovskites in particular will be the touched upon in the subsequent sections of the review. We will wrap up the review with the outlook section, where the future challenges in this field will be discussed. The pressure dependent optical properties phenomena known as piezochromism will be discussed along with the possible bridging between modeling and first principles electronic structures calculations with experimental feedback. 

\section{\textcolor{black}{Recent Experimental Progress in Hybrid Perovskites and Metal Organic Complexes}}  
As mentioned previously, a number of experimental investigations have been conducted with the focus of synthesis and characterisation of metal-organic complexes in last few years. The work of Jain and co-workers \cite{Jain1, Jain2} is one such experimental investigation of such hybrid perovskites that consists of Dimethylammonium (DMA) metal formates with the variation of metal ions as Zn, Mn, Fe, Co and Ni. They have found an exciting phase transition in dielectric constant with a hysteresis width of about 10K, and a dielectric anomaly around 160K in Dimethylammonium Zn formate, which shows antiferroelectric behaviour while cooling below 160K temperature. Additionally, a specific heat anomaly has been observed around 156K along with a order-disorder phase transition and electrical ordering, that is shown at the same temperature. The DMA cation at the centre of the ReO$_3$ type cavity is  disordered at room temperature and Mn, Co and Ni based compounds show ordering at T$_C$=8.5K, 14.9K, and 35.6K respectively with corresponding magnetic exchange coupling values of J= -0.32, -2.3, -4.85 respectively. All the three compounds exhibit canted weak ferromagnetism with a hysteresis loop below T$_C$. The AFM super-exchange and spin canting originate from non-centrosymmetric character of the atoms forming the HCOO$^-$ bridge. DMAFeF shows ferromagnetism below 20K with a corresponding phase change that has been associated with a dielectric constant anomaly. The anomaly occurred at 185K, accompanied by a hysteresis of width $\sim $10K, and a para-electric to anti-ferroelectric transition, which corresponds to structural phase transition. The corresponding T$_C$ values for Fe, Co, and Ni based hybrid perovskites are 160K, 165K and 180K respectively. The specific heat anomalies in DMAMnF can be found at 183K with the associated electrical and magnetic ordering at 8.4K. All the four compounds are multiferroic metal organic frameworks with a complex transition as compared to a straight forward three fold order-disorder model. DMA cation is found to be dynamically disordered in rhombohedral para-electric phase with a corresponding transition to monoclinic anti-ferroelectric phase that involves H bonded ordered DMA cations. The magnetic ordering can be achieved through magnetic cooling and one can observe the co-existence of antiferroelectric ordering along with with weak ferromagnetic ordering.

A combined theoretical and experimental study takes a detailed look at Dimethylammonium Cu Formate, which shows a strong one-dimensional anti-ferromagnetism. This is possible due to the Jahn Teller distortion of 3d$^9$ Cu complexes. The heat capacity shows a phase transition to a three dimensional AFM state at 5.2K. Theoretical calculations give us more insight regarding this material. It is easy to see the intra-chain spin exchange path is J$_1$, and the nearest neighbour inter-chain spin exchange paths are J$_2$ and J$_3$, where J$_1$ $\&$ J$_3$ are AFM, J$_2$ is FM and J$_1$ $\gg$ $\mid$J$_2\mid$ $>$J$_3$. The 1D AFM chains are weakly coupled through FM J$_2$ and AFM J$_3$ leading to a spin frustration between adjacent chains. A ferro-type coupling has been observed with J$_2$+J$_3$ $<$0 and the 3D magnetic structure of DMACuF shows A-AFM configuration, with Cu layers being perpendicular to the chains that are antiferromagnetically coupled in Z direction. 

There are also experimental investigations based on Neutron diffraction, which have been carried out to shed light on the order-disorder phase transition in mixed-valence iron(II)-iron(III) formate framework compound \cite{Delgado2012}. The crystal structure has been determined based on Laue diffraction data at 220 K and the subsequent refinement of the crystal structure in the low temperature phase at 45 K are extracted from the monochromatic high resolution single crystal diffractometer at 175 K. There exists experimental evidence of order-disorder phase transition associated with  dimethylammonium ion that is weakly anchored to the cavities of the organic framework. There is also a phase transition from $P3_{1/c}$ to $R\bar{3}c$, which occurs at low temperature condition due to ordering of the dimethylammonium counter-ion. It is associated with a transition from para-electric to antiferroelectric phase and it has been confirmed to be the primary reason behind the occurrence of this nuclear phase transition. It has been found that the ferrimagnetic behaviour might be due to  non-compensation of the different Fe(II) and Fe(III) magnetic moments. All these experimental studies could however be more intuitive and convincing if supported by electronic structure calculations which is essential in order to have profound understanding of the change of such intrinsic properties under external factors like temperature and hydrostatic pressure.

\section{Current Understanding of External Stimuli Driven Spin Transitions in Metal Organic Complexes}
Recently the phenomena of spin crossover has drawn a lot of attention of the materials science fraternity, which engage the communities of theoretical and experimental scientific groups, with the aim to achieve  associated  multifarious possible applications. Spin-crossover is a phenomena demonstrated in certain metal-organic complexes, that consist of transition metal ions connected by organic ligands, typically identified by a bi-stable magnetic state of the metal ion center. A unique feature of transition-metal ions binding to organic ligands is the capacity of attaining different spin states, with different magnetisation, while keeping the same valence state. Upon application of an external stimulus, such as pressure, temperature, light irradiation, or magnetic field, the transition metal ion can be reversibly switched between different spin states, for example, between a low-spin (LS) state with the smallest possible value of spin S , and a high spin (HS) magnetic state with highest possible value of S. This bi-stability of the spin state has attracted quite a bit of technological interest, as it offers the possibility of designing functional units that can be switched reversibly. Accordingly, spin-crossover (SCO) complexes are currently being considered as potential elementary units in information storage and memory devices,  spintronics devices, photo-switches,  color displays,  and, in connection with open frameworks, as gas sensors. In addition to their spin reversibility, SCO metal-organic systems are versatile and offer many possibilities for chemical functionalization, through which the desired transition property (e.g., color, transition temperature) can be efficiently tuned. Though this phenomenon, in principle, can be observed in octahedrally coordinated transition metal complexes with TM ions in d$^4$-d$^7$ electronic configurations, the most commonly observed cases are that of octahedrally coordinated iron(II) complexes with Fe$^{2+}$ ions in 3d$^6$ electronic configuration.
\textcolor{black}{SCO has been  primarily observed in organometallic molecular systems. In most cases these organometallic molecules show spin crossover mainly based on the change in geometry (primarily change in octahedral volume) which causes a competition between the crystal field splitting and Hund coupling. Thus a clear relationship of SCO with structural transitions may be derived. It is to be noted that organometallic molecules being highly susceptible to external stimuli like temperature or pressure, very easily shows such spin crossover. However not all organometallic molecules are prone to it. Spin crossover is primarily seen in $3d$ transition metals with $d^4$-$d^7$ occupancies in a octahedral ligand geometry. It is also to be noted that the magnetism in these molecular crystals are mostly collinear magnetism and involves hardly any spin orbit coupling effects.}

To have useful application in making devices, it is required to make the Spin crossover cooperative implying that we require a spin transition rather than a spin crossover, which may occur with associated hysteresis effect. The latter is of immense importance as this confers memory effect to the material. Thus much of the focus has been on the issue of cooperativity in Spin crossover. In this regard, metal organic coordination polymers or three-dimensional (3D) coordination compounds, or highly connected inorganic-organic hybrid perovskites may be considered suitable. In polymers, coordination compounds, or even perovskites the metal ion centres are linked to each other by organic bridges through which the elastic interactions and magnetic superexchange interactions may propagate very efficiently. A large number of polymeric SCO materials have been synthesized, which are found to show cooperativity at the HS-LS transition accompanied by a large hysteresis. Experiments showing SCO in hybrid perovskites are yet to be carried out though. One of the prime interests in these materials is the microscopic understanding of the phenomena of spin crossover, that is which interactions drive the cooperativity and the associated hysteresis. This understanding is of prime importance in designing potential applications of SCO materials with enhanced properties, viz. a large hysteresis effect at room temperature. There has been significant attempts by various research groups to explain the origin of this transition and its associated cooperativity. The most prevalent understanding is that the responsible factor driving cooperativity is long range elastic interactions arising due to interaction between local lattice distortions at each molecular unit. This concept however ignores completely the significance of the long range magnetic interaction that may arise between the transition metal ions via super-exchange interaction mediated through the organic ligands connecting the metal centers. The importance of magnetic super-exchange interaction has been pointed out in a density functional theory (DFT) based study which estimated the strength of magnetic exchange interaction in a Fe-triazole compound, and found it to be of the same order of magnitude as that of elastic exchange, estimated in similar compounds. There exists a few reports in which the effect of magnetic exchange interaction has been included, but the construction of the underlying Hamiltonian to study the interplay of the two has been a very recent attempt.

\textcolor{black}{It is worth mentioning here that most of the spin crossover polymers do not exist in the form of any single crystals, and are mostly either powder crystals or amorphous. A large part of the computational challenge in case of these polymeric spin crossover systems is building accurate computer models of structures representing correctly the local environment of the transition metals which mostly influence the magnetism and hence the electronic structure. The hybrid perovskites in this regard are a significant improvement upon classical spin crossover polymers since actual experimental structures are available for most of these materials. It is hence imperative to study hybrid perovskites particularly in respect of spin transitions which can be immensely useful in terms of design of energy efficient memory devices. It is to be noted that although the magnetism in a majority of the class of formate based hybrid perovskites do involve a certain amount of canting, calculations have shown that this has minimal impact and hence no large spin orbit coupling effects are seen here. }

It is to be noted here that to the best of our knowledge there is no experimental study on external stimuli driven spin transitions in hybrid perovskites as of now. We have previously theoretically predicted the existence of possible spin transition phenomena associated with hysteresis and cooperativity in Fe based hybrid perovskites as a function of pressure \cite{Banerjee2016, Banerjee2017}. We have recently shown in case of Mn based hybrid perovskites a broad hysteresis width both as a function of temperature and pressure, and this primarily brings spin transition associated with cooperativity to room temperatures which is a highly desirable feature \cite{banerjee2021}.

In the next several sections we shall expound on the various tuning methods of external stimuli driven transitions including spin transitions and other forms of transitions in both metal organic coordination polymers and hybrid perovskites driven primarily by temperature, pressure, and light irradiation.

\section{\textcolor{black}{{Tuning Paradigm of External Stimuli Driven Transitions: Effect of Temperature, Pressure and Light on Metal Organic Complexes and Hybrid Perovskites}}}
\subsection{\textcolor{black}{{Temperature Dependent Transitions in Metal-Organic Complexes}}}
Bucko and co-workers included temperature in their first principles electronic structure calculations on materials which showed spin crossover. They computed Fe(phen)$_2$(NCS)$_2$ as a single molecule and in a periodic crystal using DFT+ U approach. In addition to the ground state T = 0 K situation, they considered two different temperatures, 100 K and 500 K, using AIMD simulations. In these, the finite-temperature ionic motion is computed classically, whereas quantum mechanical calculations are employed for electronic motion, through the DFT+ U approach, and electrostatic forces connect the two motions. Computing the average coordination number, a quantity that depends on the Fe-N distances, they found a temperature-related difference that reflected a change of the average Fe-N distances to larger values at 500 K. In a earlier work, Banerjee et al have demonstrated \cite{Banerjee2014}  the temperature dependent spin crossover phenomena in Fe(II)-triazole polymer, which is shown in Fig. 2. In addition to this, ab-initio molecular dynamics investigation in this work clearly has indicated the occurrence of spin transition from a low spin LS (S=0) to a high spin HS (S=2) state with respect to the temperature increment. Moreover it has been demonstrated that at T$\sim$ 80 K, there is a clear bistability in the system, which corresponds to the hysteresis region in experiments.  The Monte Carlo aided  DFT studies have successfully demonstrated the hysteresis width is $\Delta$ T$\sim$ 20K, that is centred at T$_C$=80K and the outcome is in reasonable agreement with the experimental counterpart that have been conducted on Fe(II)-triazole polymers. 
\begin{figure}[h!]
	\begin{center}
		\includegraphics[width=\columnwidth]{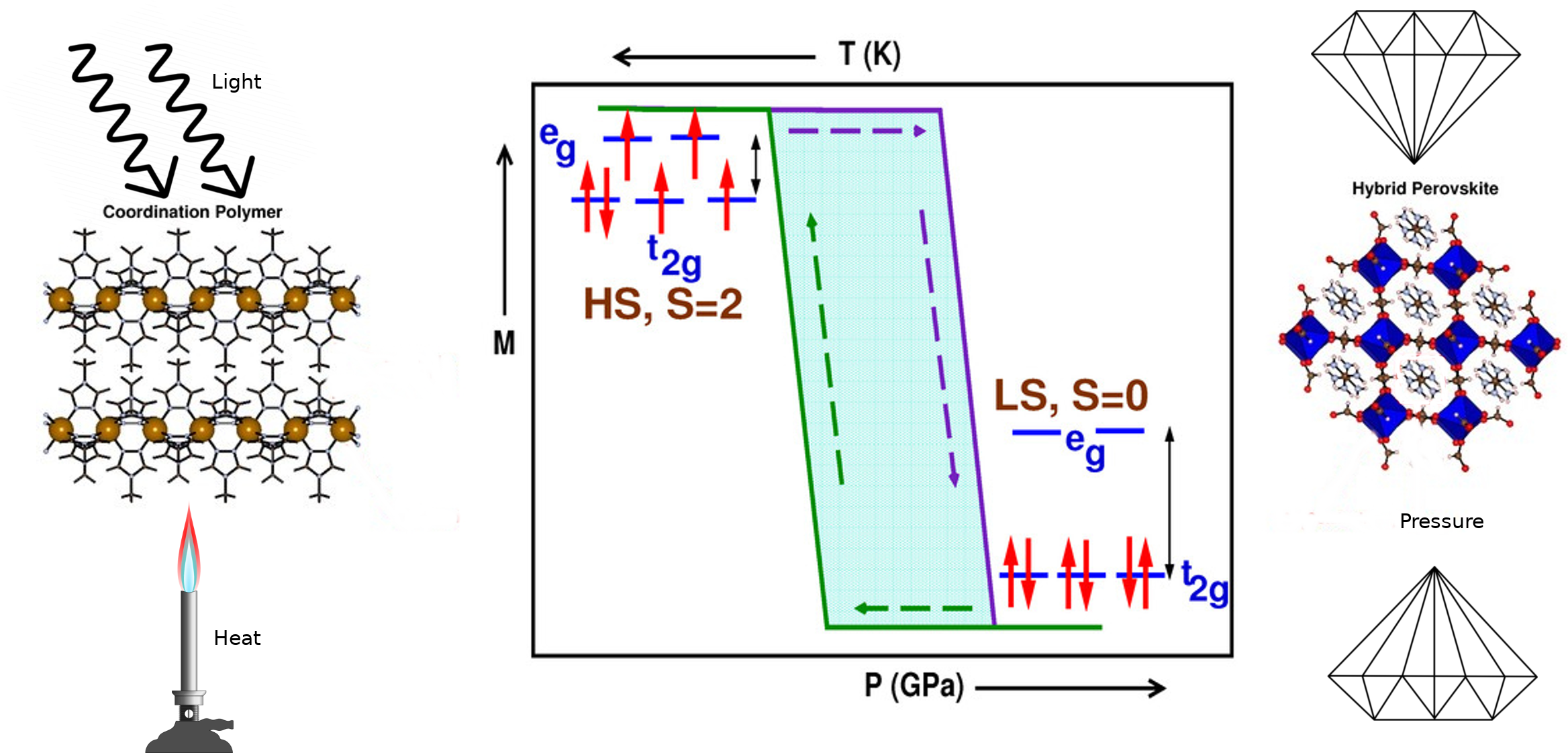}
		\caption{Figure showing temperature, pressure and light dependent Spin crossover. Temperature and pressure induced spin transitions can be driven in coordination polymers. In the middle panel computed magnetic moment per Fe atom are plotted as a function of pressure and temperature. A transition from a S=2 HS state to a S=0 LS state is seen along with a hysteresis represented by the cyan shaded region. The electronic configurations in the HS and LS states and the corresponding electron distribution between $e_g$ and $t_{2g}$ states is also shown. The leftmost panel shows that the coordination polymers may be subjected to heat or light irradiation to show spin crossover. The rightmost panel shows the application of pressure to hybrid perovskites to give rise in transition may be easily achieved by application of a diamond anvil cell in experiments, and though it has been studied primarily theoretically, may be a fruitful experimental study. }
	\end{center}
\end{figure}
\begin{figure}[h!]
	\begin{center}
		\includegraphics[width=\columnwidth]{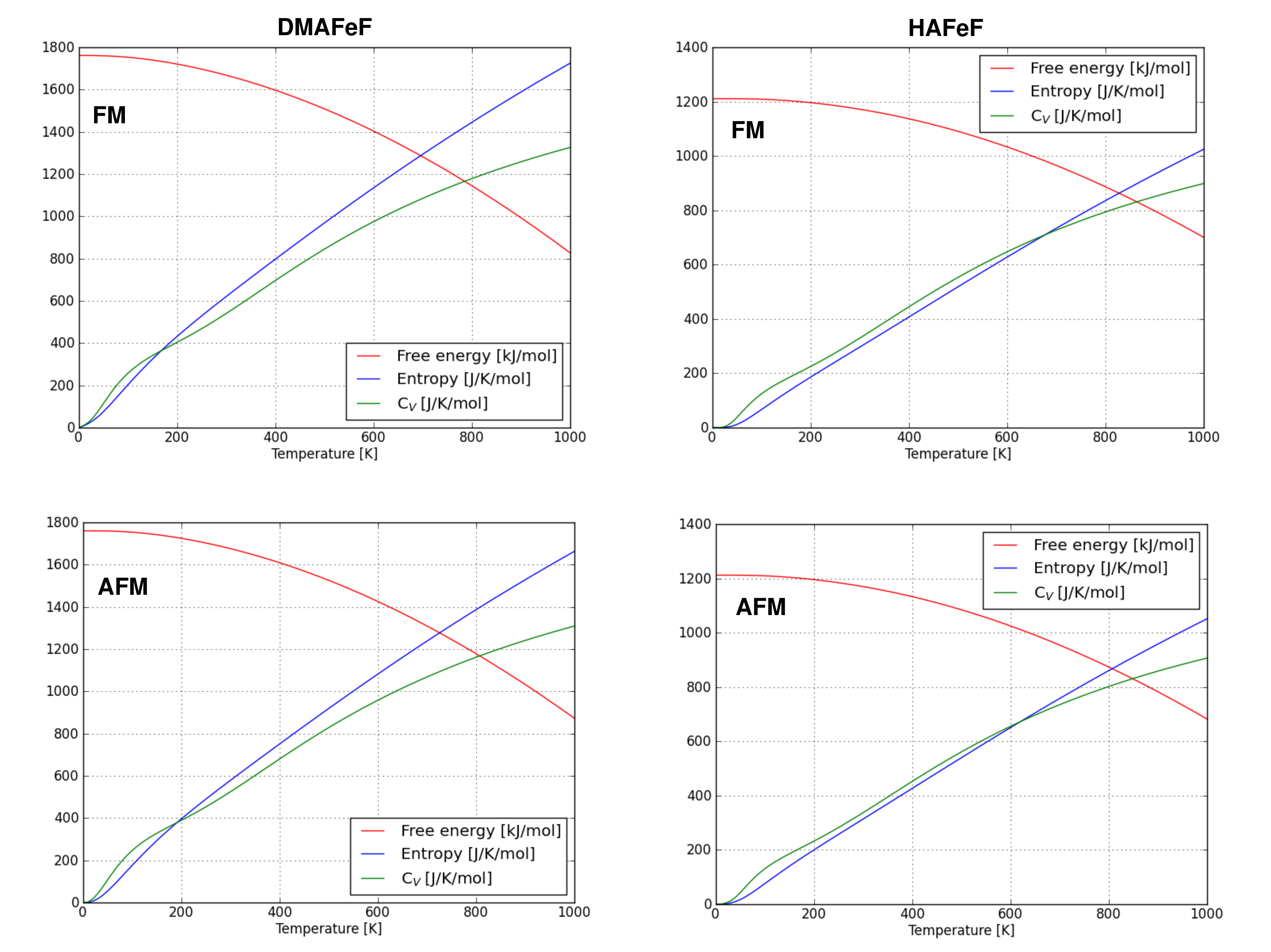}
		\caption{Figure showing phonon properties in Metal Organic Complexes linking magnetism with temperature dependence.}
	\end{center}
\end{figure}
A two-step SCO transition \cite{sudipto-sco1,sudipto-sco2,tsd-rev},  has also been found in a tetranuclear Fe(II) molecular system, Fe(II)$_4$($\mu$-CN)$_4$(bpy)$_4$(tpa)$_2$ ](PF$_6$)$_4$ (where bpy = 2,2'-bipyridine and tpa = tris(2-pyridylmethyl)amine). In the molecular unit, there are four in-equivalent, octahedrally coordinated Fe centers in a square planar arrangement. The molecular units crystallize in a triclinic unit cell in which the ferrous squares are van der Waals bonded and separated by (PF$_6$)$_4$ - counter ions. Upon heating, there is a first LS-HS-1 SCO transition that has been observed at T$_1$ $\sim$ 160 $\pm$ 20 K, when one of the Fe atoms transforms to HS S = 2. Further increase of the temperature induces a second SCO transition at about T$_2$ $\sim$ 370 $\pm$ 50 K, where a second Fe transforms to HS, leading to HS-2. Electronic structure calculations within the framework of first-principles DFT+ U, in combination with AIMD have been performed to simulate the temperature-dependent spin state of this large system containing 400 atoms in the unit cell. Temperature has been varied step-wise between 0 K and 500 K, which shows a first SCO conversion at 138($\pm$10)K and a second one at 375($\pm$20)K, which are in good agreement with experiments. The Fe centers coordinated with six nitrogen atoms undergoes the LS-HS transition, whereas those which were coordinated by four nitrogen and two carbon atoms, remained in LS. The calculated temperature evolution of the average bond lengths at Fe sites is comparable to experimental studies. At the SCO transitions, bond length elongation from less than 2$A^0$ to $\sim$ 2.18$A^0$ has been predicted, that is consistent with the well-known relation between spin state and bond length. The tetranuclear unit forms an open structure, which enables the incorporation of guest molecules like CO$_2$ , and H$_2$O in the ferrous square. Exploring the possibility of chemical switching, it was found that inclusion of a guest molecule in the system, slightly increases the bond lengths and could thereby induce an LS–HS-1 transition at already low temperatures and a HS-1--HS-2 transition at moderate temperatures ( $\sim$ 250 K). The possibility of such chemical switching would certainly add a further dimensionality to these exciting materials. 

\subsection{\textcolor{black}{{Temperature Dependent Transitions in Hybrid Perovskites}}}

\begin{figure}[h!]
	\begin{center}
		\includegraphics[width=\columnwidth]{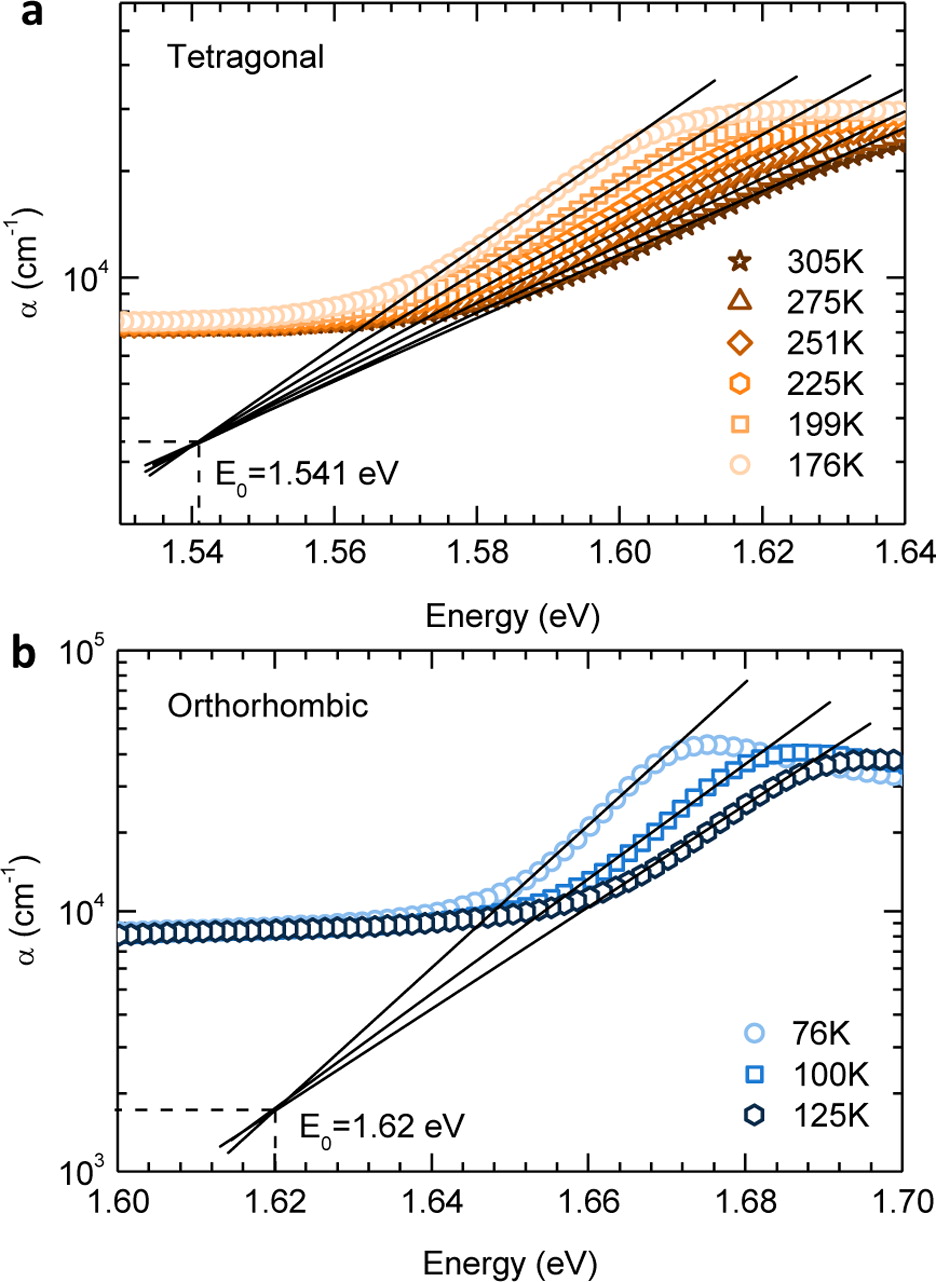}
		\caption{(a) Tetragonal and (b) orthorhombic phase of \ce{MAPbI3} showing variation of photon energy with the logarithmic absorption coefficient($\alpha$) at various temperatures. Reprinted with permission from ref \citenum{Shivam}. Copyright 2016 American Chemical Society.}
 \label{temp} 
	\end{center}
\end{figure}

Organic-Inorganic hybrid perovskites have a high power conversion efficiency but they are very unstable at high temperatures and moisture. This is the reason for their hindrance in commercialisation as solar cells. Understanding the degradation of perovskite solar cells at high temperatures can help us overcome these shortcomings. In this section, we traverse through the degradation process of various perovskite solar cells under the influence of temperature as an external stimulus. The effect of temperature on methylammonium lead iodide (\ce{MAPbI_3}) was investigated computationally using molecular dynamics simulations. A recent work \cite{carignano2014thermal} simulated the tetragonal structure of \ce{MAPbI3}. With change in temperature, the HOMO-LUMO energy gap fluctuated at its central value but there was smoothness around the edges of projected density of states (PDOS) \cite{carignano2014thermal}. The change in optical properties of \ce{MAPbI_3} with  temperature was reported experimentally. It was reported in this work \cite{Shivam} that the bandgap decreased with the decrease in the temperature\cite{Shivam} for both tetragonal(T $>$ 163 K) and orthorhombic (T $<$ 163 K) phases of \ce{MAPbI_3}(Figure \ref{temp}), although for the crystalline semiconductors\cite{Olgui2002}\cite{Varshni1967TemperatureDO}, an opposite trend was observed. The optical absorption spectra showed a strong excitonic effect \cite{Innocenzo2014, Docampo2013} at low temperatures. The full width at half maxima (FWHM) of the excitonic peak also varied with the temperature. Till 60 K, it decreased in a linear fashion. Flattening of the curve was reported at lower temperatures. Through the phase transition, it again started decreasing \cite{Innocenzo2014, Docampo2013}

\textcolor{black}{At temperatures between \SI{85}{\degreeCelsius} to \SI{120}{\degreeCelsius} without moisture \ce{MAPbI3} was reported to be intrinsically stable\cite{Park}. However after encapsulation including the spiro-MeOTAD, degradation was observed at a temperature of \SI{85}{\degreeCelsius}. With increase in  temperature, charge separation from \ce{MAPbI3} to spiro-MeOTAD was inhibitated\cite{Park}. After aging of the device at \SI{85}{\degreeCelsius}, the spiro-MeOTAD layer was replaced with a new one. It was observed that the efficiency of the device went back to its original value. However for an inverted device where \ce{NiO} was used as a hole transport material, after an increase in temperature the efficiency remained at 74\% of the original PCE. This showed that selective contacts can help the device retain its efficiency at high temperatures\cite{Park}. Thermal aging of a device made up of \ce{MAPbI_3} at \SI{85}{\degreeCelsius} for 500h was shown \cite{Liyuan}. It was observed that the PCE was retained upto 80\% of the original efficiency of 19.19\%\cite{Liyuan}. This was possible due to additive engineering of the device. In a recent work \cite{Meng2021}, the thermal stability in \ce{(FAPbI_3)_{1-x}MAPb(Br_{3-y}Cl_y)_x} organic-inorganic perovskite with mixed cation was reported\cite{Meng2021}. Piezochromism was observed with increase in temperature on these perovskite films from \SI{25}{\degreeCelsius} to \SI{250}{\degreeCelsius}. The change in colour was from black to yellow. There was linear reduction in PCE with increase in temperature. From the XRD pattern it was evident that at high temperatures, the perovskite decomposed to \ce{PbI_2}. From the UV-vis absorption spectra, it was observed that as temperature is increased, the bandgap reduces \cite{Meng2021} except at \SI{250}{\degreeCelsius}. The bandgaps at temperatures \SI{85}{\degreeCelsius}, \SI{100}{\degreeCelsius}, \SI{150}{\degreeCelsius}, \SI{200}{\degreeCelsius}, and \SI{250}{\degreeCelsius} was found to be \SI{1.566}{\electronvolt}, \SI{1.564}{\electronvolt}, \SI{1.543}{\electronvolt}, \SI{1.508}{\electronvolt}, and \SI{2.378}{\electronvolt}, respectively. The bandgap of the pristine system is \SI{1.569}{\electronvolt}. However when the device was heated for 10 minutes at \SI{85}{\degreeCelsius}, efficiency increased by 15\%. During this heating process,  \ce{PbI2} passivisation occurred which in turn increased efficiency\cite{Meng2021}.}

In context of temperature dependent tuning of formate based hybrid perovskites a relationship between phonon modes and corresponding magnetic states may be observed in hybrid perovskites. Even though a direct observation of temperature driven magnetic switching has not yet been materialised in formate based hybrid perovskites,  we show here in Fig. 3 the correlation between a possible temperature driven spin switching between FM and AFM states, and phonons in case of DMAFeF and HAFeF. Banerjee et al. have recently shown the possibility of temperature driven spin transitions in DMAMnF - a Mn based formate hybrid perovskite employing ab-initio MD calculations which brings the temperature driven spin transition associated with hysteresis to near room temperatures making it highly accessible for device applications. \cite{banerjee2021}

\subsection{\textcolor{black}{{Pressure Dependent Transitions in Metal-Organic Complexes}}}

Temperature and pressure-dependent AIMD simulations have been performed on considerably large metal organic systems. One such system is a recently discovered Fe-Nb bimetallic framework, Fe$_2$[Nb(CN)$_8$].(4-pyridinealdoxime)$_8$.2H$_2$O, which forms a 3D connected network of cyanide-bridged Fe-Nb atoms with the unit cell containing 290 atoms. This material has been reported to display both a temperature- and light-induced SCO transition. In the LS state, the S = 0 Fe(II) centers are connected via a Nb(IV) ion in S = 1/2, whereas in the HS state above about 130 K, the S = 2 spins on Fe ions are coupled via an antiparallel S = 1/2 spin on the interconnecting Nb. There are reports in this Fe-based framework, employing a combination of DFT+ U and AIMD simulation, to sample stepwise pressure-temperature phase space. The experimentally observed temperature driven SCO transition in the Fe variant was accurately described. Moreover, novel spin-state transitions were predicted that could only be reached via specific pressure-temperature combinations. One of these was the novel intermediate spin (IS) state S = 1 appearing on the Fe atoms at modest pressures \cite{Bressler489} before the LS state was reached at a higher pressure. The sharp SCO transitions are associated with a hysteresis depending on the increase or decrease in pressure, indicating the cooperative nature of the transitions. The Fe S = 1 and Nb S = 1/2 spins were aligned in parallel in the IS state of the material. Astonishingly, a further new HS state (HS-2) was achieved by applying both pressure and temperature to this IS state, where long-ranged ordering was observed similar to HS-1, but with a parallel alignment of Fe S = 2 spins and Nb S = 1/2 spins. Further enhancement of the pressure turned the material to LS state system. One of the crucial parameters identified to trigger SCO transitions with temperature or pressure is bond length as it was found in previous reports as well. The average Fe-N bond length along the cyanide bridges is reduced from 2.04 $A^0$ at the HS-1 state to 1.93 $A^0$ at the IS state and 1.91$A^0$ at the LS state. Metal-organic materials that contain several SCO centers can display more  such complex behaviour, which needs further extensive investigation both from theory and experiments.

\begin{figure}[h!]
	\begin{center}
		\includegraphics[width=\columnwidth]{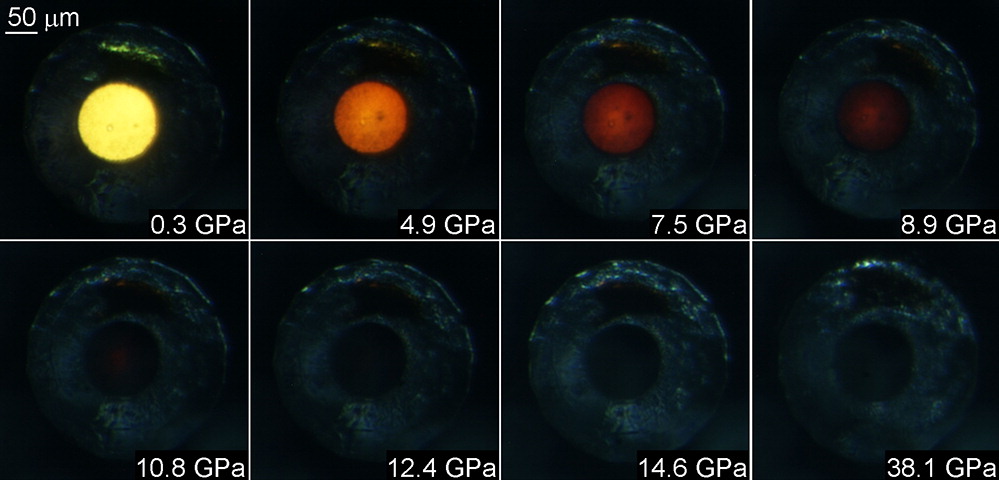}
		\caption{Piezochromic effect in \ce{(EDBE)[CuCl4]} showing transistion from translucent yellow to opaque black with the increase in the pressure.Reprinted with permission from ref \citenum{Adam}. Copyright 2015 American Chemical Society. }
	\end{center}
	\label{Adam}
\end{figure}

\begin{figure}[h!]
	\begin{center}
		\includegraphics[width=\columnwidth]{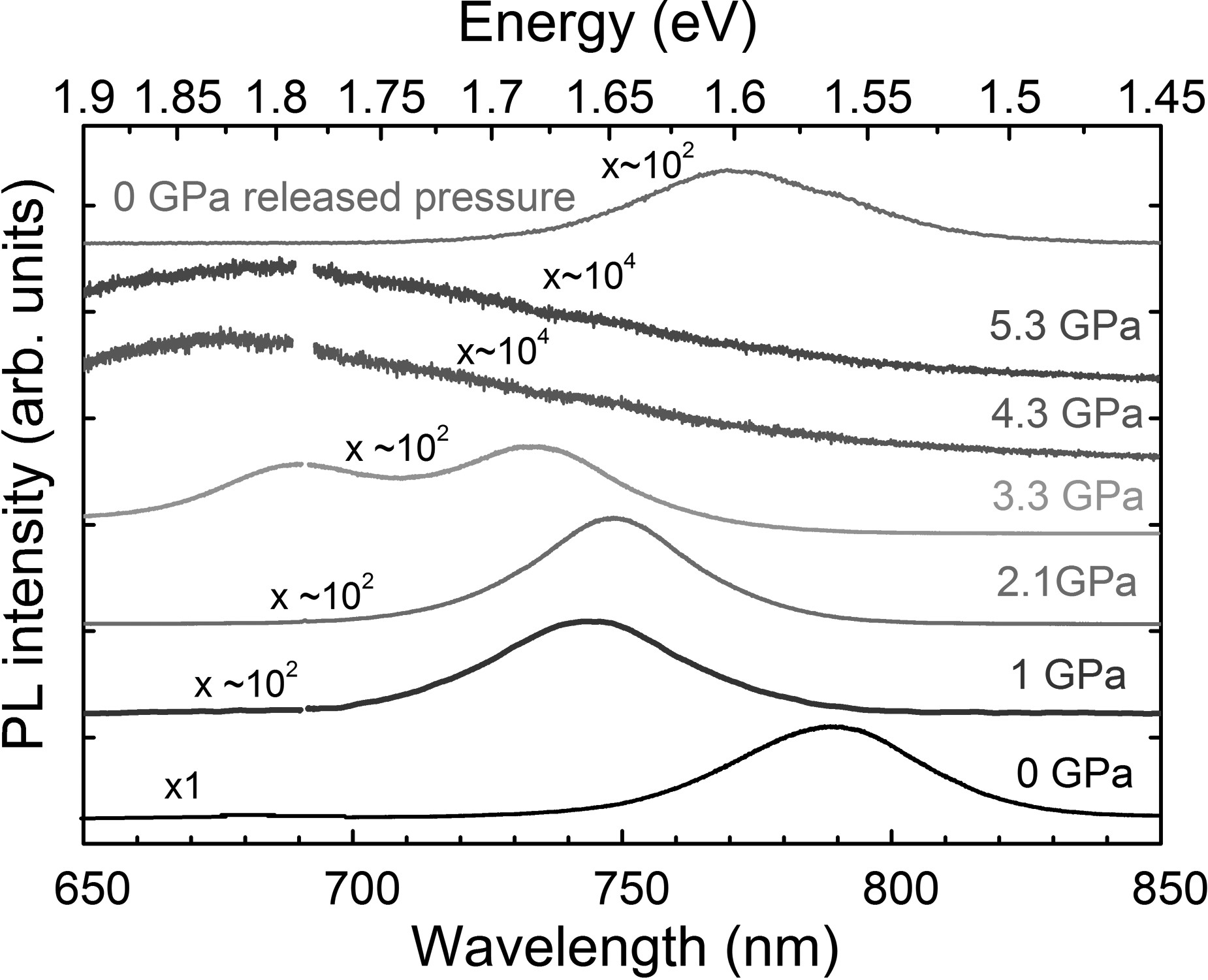}
		\caption{Variation of photoluminescence spectra of \ce{MAPbI3} under increase in pressure on the crystal.Reprinted with permission from ref \citenum{Capitani}. Copyright 2016 American Institute of Physics.}
	\end{center}
	\label{Capitani}
\end{figure}
\subsection{Pressure Dependent Transitions in Hybrid Perovskites}
 Pressure dependent spin crossover was recently investigated thoroughly in hybrid perovskites as shown in Fig. 2, for the first time from DFT based electronic structure calculations \cite{Banerjee2016}.  By considering two materials, Dimethylammonium Iron Formate (DMAFeF) and Hydroxylamine Iron Formate (HAFeF),  it was demonstrated that upon application of external pressure, a spin state transition at two different pressures occur. This is strongly dependent on the mechanical strength of the materials, which in turn is highly influenced by the hydrogen bonding in the system. HAFeF having stronger H bonding has a higher bulk modulus compared to DMAFeF and hence a higher value of critical pressure for spin transition. While releasing pressure, it was observed that both the perovskite systems show hysteresis associated with the spin transition where HAFeF has a much greater hysteresis width as compared to that of DMAFeF.

It was found that on exerting pressure (P) in the range of $\sim$ 5-7 GPa causes a drastic change in the electronic structure of DMAFeF and HAFeF. Interestingly, the ground states turned out to be non-spin polarized in both the compounds with 0 $\mu_B$ associated magnetic moment. This confirms a spin-state transition from HS (S = 2) state to LS (S = 0) state, obtained by external hydrostatic pressure. In the next step, in order to find out the critical pressure corresponding to the spin-state transition for the two compounds, the pressure was increased in stepwise manner (0.6-0.7 GPa as the step size), starting from the ambient pressure condition. The plot of magnetic moment (M) versus pressure (P), shows a spin-state change from HS with total magnetic moment of 4 $\mu_B$/Fe to LS with a total moment of 0 $\mu_B$/Fe at pressure ($P_{C\uparrow}$) of 4.7 GPa for DMAFeF and 6.6 GPa for HAFeF. This may be attributed to the strong influence of choice of the organic cation on the required optimal pressure for spin-state transition. It was observed that the optimal pressure, which was required for the transition from LS with a total moment of 0 $\mu_B$/Fe to HS with total magnetic moment of 4 $\mu_B$/Fe occurred at a different pressure ($P_{C\downarrow}$) compared to $P_{C\uparrow}$, having values of 2.5 GPa for DMAFeF and 1.4 GPa for HAFeF. There is significant hysteresis effect in M-P data in both the compounds, associated with the hysteresis width of  2.2 GPa for DMAFeFe as compared to 5.2 GPa for HAFeF, with the former being more than a factor of 2 smaller than the latter. This lead to the conclusion that spin-switching properties under external hydrostatic pressure are strongly dependent on the choice of A site cations in hybrid perovskites. This constitutes a key finding of this particular study where not only spin crossover was demonstrated for the first time in such complexes but also the mechanism to chemically tune electronic properties was explored within the framework of DFT.

In the high pressure LS state a large band gap of $\sim$1 eV opens up between the fully occupied Fe t$_{2g}$ states and completely empty Fe $e_g$ states. The spin-state transition thus may be accompanied by a significant change in the overall band gap, which may manifest in corresponding change in optical response and change in colour. There has been a recent few experimental works, where band gap closure and therefore the semiconductor to metal transition was shown to be possible using diamond anvil pressure. This phenomena is known as piezochromism \cite{Jaffe2015, Umeyama2016, Kojima2009, Giovanni2015, Savory2016, Swainson2007, arnab-piezo, arnab-piezo-2}, which might be able to produce transparent complexes routinely that can also be used as a substitute of transparent conducting oxide (TCO) for the transparent window based applications. Exerting large hydrostatic pressure by using diamond anvil cell, it may be possible achieve an irreversible structural phase transformation, whereby the system may remain in the high pressure phase and show metallicity forever and therefore can be used as a replacement of TCO for transparent building technology \cite{TCO}. This can open up a new area of research  which is important both from the perspective of fundamental understanding as well as in terms of corresponding industrial applications.

\textcolor{black}{Next we explore the effects of applying pressure on various different hybrid perovskite systems. The main outcome of compression and decompression was piezochromism which had a direct link with the change in bandgap. Hence, applying pressure as an external stimulus helps in tuning the bandgap which in turn makes the system suitable for various photovoltaic applications.}

\textcolor{black}{For \ce{MASnI3}, phase transitions occurring due to application of pressure was reported to be $Pm\bar3m$ to $Im\bar3$ to $Immm$, whereas for \ce{FASnI3} and \ce{M_{0.5}F_{0.5}SnI3} was from $Pm\bar3m$ to $Im\bar3$ to $I4/mmm$ [MA or Methylammonium is \ce{CH3NH3^+} ,and FA or Formamidinium is \ce{NH2CH\bond{=}NH2^+}]. At low pressure(\SI{0.5}{\giga\pascal}), continuous second order phase transition was observed from primitive to body centered cubic\cite{Lee}. This was due to straining of Sn-I bonds.  At higher pressure(\SI{2.2}{\giga\pascal}, and \SI{1.9}{\giga\pascal}), due to tilting and distortion of the octahedra \ce{SnI6}, a phase transition occurred to orthorhombic and tetragonal phase. When decompression was done on the samples after amorphisation, reversibility in the phase was observed. Templating effect of the organic cation was the reason for this reversibility. The overall bulk moduli for \ce{MASnI3}, \ce{FASnI3}, and \ce{MA_{0.5}FA_{0.5}SnI3} was reported to be  12.6, 8.0, and 11.5 \si{\giga\pascal}. This showed that these systems were highly compressible on application of pressure\cite{Lee}. The structural changes on the perovskite system \ce{EA2CuCl4} under the effect of hydrostatic pressure was observed\cite{Kenji}, where EA is ethyl-ammonium. A phase transition from orthorhombic to monoclinic phase was observed at around \SI{4}{\giga\pascal}. $P2_1/a$ was the space group at this pressure  which is similar to that of \ce{MA2CuCl4}. Jahn Teller distortion was also subdued below \SI{4}{\giga\pascal}. Piezochromism was also observed under pressure. The colour of the system changed from yellow to red with increase in pressure. Above this high pressure, the RXS intensity also becomes null. It was inferred that the ionic environment of the \ce{Cu^{2+}} ions had changed\cite{Kenji}. }

\textcolor{black}{ Pressure effect on \ce{MAPbBr3} was studied and associated phase transition\cite{Swainson} of the crystal was seen from $Pm\bar3m$ to $Im\bar3$ below \SI{1}{\giga\pascal}. The amorphisation of the crystal occurred at \SI{2.8}{\giga\pascal}. From the DFT calculations, it was observed that the volume of the crystal lowered on applying pressure due to the tilting of the \ce{PbBr6} octahedra rather than the shrinkage phenomena. This reduction in the volume leads to a change in the phase transition as the energy gain was low for orientational ordering.\cite{Swainson}
For the first time room temperature electrical conductivity\cite{Adam} of layered C-Cl perovskite system was observed in \ce{(EDBE)[CuCl4]}. Piezochromism was observed on applying pressure on the crystal as seen in Figure \ref{Adam}. The colour of the crystal changed from translucent yellow to red to opaque black on application of pressure of magnitude \SI{60}{\giga\pascal}. At \SI{4}{\giga\pascal}, the phase transition from yellow to red occurred. Another phase transition occurred at around 7 and \SI{9}{\giga\pascal}. At \SI{51.4}{\giga\pascal}, where the conductivity was reported to be $2.9 \times 10^{-4}$ S/cm. It was observed to be 5 times greater than the conductivity reported at the ambient pressure. The bandgaps at \SI{39.7}{\giga\pascal} was found to be \SI{1}{\electronvolt} and at\SI{40.2}{\giga\pascal}, the activation energy was found to be  0.232 \si{\electronvolt}. This lead to an electronic conduction pathway owing to the $d$ orbitals.\cite{Adam} Application of hydrostatic pressure\cite{Wang} of \SI{38}{\giga\pascal} on \ce{CH3NH3PbBr3} to tune the photocurrent properties and bandgap was shown in this work\cite{Wang}. At ambient pressure, the space group of \ce{CH3NH3PbBr3} was reported to be $Pm\bar 3m$. Between 0.4 and \SI{1.1}{\giga\pascal}, the intermediate space group was found to be cubic $Im\bar3$. From 1.8 to 4 \si{\giga\pascal}, the space group changed to $Pnma$ till complete amorphisation. From first principles calculations, it was observed that there was a decrease in the bandgap when hydrostatic pressure was increased. An increase in  bandgap was observed due to amorphisation due to pressure effect. This was because of breakdown of long range order which suppressed the overlap of the elemental orbital. From the DOS it was observed that the primary contribution to valence band was from Br $4p$ orbitals, whereas the major contribution to conduction band was from Pb $6p$ orbital.} Upon increase of pressure, the \ce{PbBr6} octahedra shrank leading to a decrease in the bandgap\cite{Wang}.
 
 \textcolor{black}{The space group of the crystal at room temperature was reported to be $I4/mcm$. At \SI{0.26}{\giga\pascal}, the space group changed to orthorhombic i.e. $Imm2$. The transition to the amorphous state of the crystal started at \SI{2.3}{\giga\pascal} and was completed at \SI{3}{\giga\pascal}. At \SI{1}{\giga\pascal}, the peak of the PL spectra(Figure \ref{Capitani}) was observed at \SI{1.66}{\electronvolt}. At \SI{2.1}{\giga\pascal}, the peak was at \SI{1.66}{\electronvolt}. The peak went back to \SI{1.685}{\electronvolt} at a pressure of \SI{3.3}{\giga\pascal}. But, a new peak at \SI{1.80}{\electronvolt} was observed at this pressure. On decompression of the crystal, the PL spectra were found to be reversible. Piezochromism from black to red colour was also observed with an increase in pressure\cite{Capitani}.  Application of pressure on \ce{MAPbI3} nanorods was studied in this work\cite{Tianji}. Amorphization was observed at \SI{20}{\giga\pascal}. This high value of pressure might be due to the absence of transmitting medium\cite{Tianji}. Phase transition from orthorhombic to tetragonal phase of commercial sample\cite{wang2015pressureinduced} of \ce{MAPbI3} at \SI{0.3}{\giga\pascal} was reported, although the amorphization of the sample was said to be at \SI{7}{\giga\pascal} pressure. Zou et al. studied application of pressure as an external stimulus on the crystal \ce{FAPbBr3}\cite{Zou}. At ambient pressure, the bandgap of \ce{FAPbBr3} was reported to be \SI{2.23}{\electronvolt} and the colour was orange with transparency in the crystal. The colour of the crystal changed to red at about \SI{2.2}{\giga\pascal} pressure\cite{Zou}. On application of further pressure, the colour changed to yellow and then to colourless for pressures above \SI{4.1}{\giga\pascal}. Upon decompression, the colour of the sample came back to its original colour. Thus \ce{FAPbBr3} showed piezochromism on application of pressure. From the absorption spectra, it was observed that there was a change in the bandgap on the application of pressure\cite{Zou}. At 0.6 and \SI{2.2}{\giga\pascal} pressures an abrupt change in the bandgap was reported. The space group of \ce{FAPbBr3} at ambient pressure was $Pm\bar3m$ which changed to $Im\bar3$ at \SI{0.53}{\giga\pascal}. At \SI{2.2}{\giga\pascal} pressure, the space group changed from $Im\bar3$ to $Pnma$. Amorphization of the crystal structure was observed at pressure \SI{4.0}{\giga\pascal}. When this phase transition of \ce{FAPbBr3} is compared with that of \ce{MAPbBr3}, it was observed that \ce{FAPbBr3} was less compressible on the application of pressure than latter\cite{Zou}.} 

\subsection{\textcolor{black}{Pressure Dependent Transitions in Inorganic Perovskites}}
In this context of pressure driven tunability of hybrid perovskites it may also be interesting to look at an extreme case where instead of an organic moiety one deals with a large alkali metal ion, which may also be susceptible to pressure. This has been studied recently in the halide perovskite family. Yurong et al. reported the effects of pressure on the \ce{CsPbX3} (X = I, Br, Cl) halide perovskite systems by employing first principles DFT method \cite{Yurong}. Phase transitions in the order of $non-Pv-Pnma$, $C2/m-I$(\SI{11}{\giga\pascal}), and $C2/m-II$(\SI{52}{\giga\pascal}) was observed for \ce{CsPbI3} on applying pressure in the range of 0 - 120 \si{\giga\pascal}. There was a decrease in the cell volume and shortening of the polyhedral bonds. Whereas for this pressure range, the phase transition for  \ce{CsPbBr3} was in the order $Pv-Pnma$, $non-Pv-Pnma$(\SI{0.35}{\giga\pascal}), $C2/m-I$(\SI{9}{\giga\pascal}), and $Cmcm$(\SI{85}{\giga\pascal}). For \ce{CsPbCl3}, the order of phase change was found to be same as \ce{CsPbBr3}, but at different values of pressure\cite{Yurong}. $non-Pv-Pnma$ space group was attained at \SI{1.5}{\giga\pascal}, $C2/m-I$ at \SI{12}{\giga\pascal}, and $Cmcm$ at \SI{80}{\giga\pascal}. 
The bandgap of \ce{CsPbI3} at ambient conditions was reported to be \SI{2.5}{\electronvolt}(indirect). At a pressure of \SI{9}{\giga\pascal}, the bandgap reduced to \SI{1.75}{\electronvolt}. At \SI{20}{\giga\pascal}, the bandgap further reduced to \SI{1.4}{\electronvolt}. Metallisation of the structure occurred at \SI{65}{\giga\pascal}. Piezochromism phenomena in \ce{CsPbI3} was also very large with the value of \SI{0.058}{\electronvolt\giga\pascal^{-1}}. The bandgap of \ce{CsPbBr3} at ambient pressure was reported to be \SI{2.09}{\electronvolt}(direct). At pressure of \SI{0.3}{\giga\pascal} the bandgap increased from 2.07 to 2.87 \si{\electronvolt}\cite{Yurong}. On further application of pressure the bandgap started decreasing again. With increase in pressure from 0 - \SI{1.5}{\giga\pascal}, the bandgap reduced from 2.5 to \SI{2.4}{\electronvolt} for \ce{CsPbCl3}. At \SI{1.5}{\giga\pascal} the bandgap increased from 2.4 to \SI{3.3}{\electronvolt}. With further increase of pressure, the bandgap was reported to decrease again\cite{Yurong}. Importantly piezochromism in lead free inorganic perovksites have also been studied extensively in DFT literature \cite{arnab-piezo, arnab-piezo-2}. Importance of spin orbit coupling and Rashba SOC effects have also been noted in ab initio studies \citenum{Chakraborty2021, Kaur2022}.

\subsection{Pressure Dependent Transitions in 2D hybrid Perovskites}

The 2D inorganic perovskites\cite{Guloy, Dohner} are an emerging class of perovskite systems with quantum confinement effects and and better charge transport properties\cite{Ballouli}. They show stability and are more prone to pressure effects\cite{Xujie} due to soft lattices when compared to the 3D perovskites. In this section we will see how the bandgap can be tuned by application of pressure for the 2D layered perovskite systems\cite{Manasa, Mohanty_2022}.\\ 
Ren et al. tuned the bandgap of 2D perovskite\cite{Ren} \ce{(PEA)2PbBr4} by applying pressure on the system. With increase in pressure, red shift was observed in the PL spectra. A decrease in the bandgap was observed in the pressure range of \SI{0}{\giga\pascal}-\SI{10}{\giga\pascal}, and from \SI{28.4}{\giga\pascal}-\SI{48.2}{\giga\pascal}. This was due to the lattice contraction of the system. Whereas from \SI{10}{\giga\pascal}-\SI{28.4}{\giga\pascal}, there was increase in the bandgap with increase in the pressure. This was due to pressure induced amorphisation.
In another work, \SI{10}{\giga\pascal} pressure was applied on butylammonium lead halide\cite{Yin} \ce{(C4H9NH3)2PbI4}. By application of \SI{0.1}{\giga\pascal} pressure, Pb - I bond length increased, whereas Pb - I - Pb bond angle decreased. However above \SI{1.4}{\giga\pascal}, opposite effect was observed for the Pb - I bond length and Pb - I - Pb bond angle. The bandgap reduced to \SI{2.03}{\electronvolt} at \SI{5.3}{\giga\pascal}. Liu et al. tuned the optical and structural properties\cite{Richard} of 2D Ruddlesden–Popper hybrid perovskite \ce{(BA)2(MA)Pb2I7}, where BA is butylammonium. For pressure till \SI{4}{\giga\pascal} red shift was observed whereas till \SI{13}{\giga\pascal} blue shift was observed. This change from red shift to blue shift at low pressure range was due to atomic distortions. However at \SI{13}{\giga\pascal}, again red shift was observed. This atomic distortion was observed due to layer-to-layer compression occuring for low pressure, but interlayer compression occurring for high pressure.
\ce{MA3Bi2Br9} is another Bi-based 2D perovskite in which pressure dependent structural and optical properties were observed\cite{Quan}. At ambient pressure \ce{MA3Bi2Br9} crystallised to trigonal $P\bar{3}m1$ space group. But at a pressure of \SI{5}{\giga\pascal}, the space group changed to monoclinic $P2_1/a$. Bandgap was also tuned with increase in pressure. Red shift was observed till pressure of \SI{4.1}{\giga\pascal}. From \SI{4.6}{\giga\pascal} to \SI{5.5}{\giga\pascal}, blue shift was observed. This shift in the bandgap evolution was due to the tilting and distortion of the octahedra. Till \SI{10.4}{\giga\pascal}, again red shift was exhibited.
Mario et al synthesised a novel 2D layered perovskite\cite{Mario} \ce{(4BrPhMA)2PbBr4} and \ce{(4BrPhA)6Pb3Br12} (\ce{4BrPhMA}: (4-bromophenyl) methylammonium, \ce{4BrPhA}: (4-bromophenyl)ammonium). On application of small pressure of \SI{10}{\kilo\bar} a red shift was observed in case of \ce{(4BrPhMA)2PbBr4}. The inorganic layer also degraded with the application of further pressure. However blue shift was observed at \SI{10}{\kilo\bar} for \ce{(4BrPhA)6Pb3Br12}. \ce{DA2PbI4} and \ce{DA2GeI4} (DA = decylammonium) 2D perovskites\cite{Marta} were recently investigated by applying pressure till \SI{12}{\giga\pascal}. At \SI{11.5}{\giga\pascal} red shift was observed for \ce{DA2PbI4}. There is a phase transition from $Pbca$ to $P2_1/a$ at mere \SI{0.36}{\giga\pascal}. However for \ce{DA2GeI4}, at \SI{0.2}{\giga\pascal} sudden blue shift was observed, but again there is a red shift observed with further increase in pressure.

In the context of tunability of properties in 2D hybrid perovskites it is worth noting that driving ferromagnetism in 2D hybrid perovskites may be a novel route of tunability. Atomically thin films of inorganic-organic perovskite materials such as CH$_3$NH$_3$PbI$_3$ have been synthesised which are promising solar cell candidates. Dou et al. \cite{Dou2015} showed that thin films - a single unit cell or a few unit cells thick - of a related composition, (C$_4$H$_9$NH$_3$)$_2$PbBr$_4$, form squares with edges several micrometers long, and exhibit strong and tunable blue photoluminescence. Many such layered materials are well eligible for exfoliation of 2D thin films. Of particular interest in this context are ones containing transition metal ions such as Cu$^{2+}$, Ni$^{2+}$, Mn$^{2+}$ or Fe$^{2+}$ which opens up the prospect of magnetism \cite{Burschka2013, Lee2012}. Of most interest among these in context of 2D ferromagnetism are those of Cu$^{2+}$ based systems, which in their bulk form show ferromagnetic ordering in the inorganic part of the compounds.

 A recent computational study provides a novel way of tuning 2D ferromagnetism in hybrid perovskites by exfoliating 2D layers from existing 3D hybrid perovskites \cite{Nafday2019}. Taking the example of four of these Cu$^{2+}$-based layered organic-inorganic complexes of a general formula (R-NH$_3$)$_2$CuCl$_4$ with R=CH$_3$CH$_2$(EA), C$_3$H$_5$CH$_2$ (CPMA), C$_6$H$_5$CH$_2$CH$_2$(PEA), and C$_10$H$_7$CH$_2$ (Naphtha-4-MA) \cite{Polyakov2012}, Nafday et al explores the properties of the 2D monolayers exfoliated layered 3D hybrid perovskites through ab-initio density functional theory (DFT) combined with a model Hamiltonian. Their study shows that  that 2D ferromagnetism may be stabilized in a monolayer of a corner-shared network of CuCl$_6$ octahedra of these compounds. They show that the computed cleavage energies are a factor of 2-3 smaller than that of graphite, hence opening up a route to making experimental synthesis of such monolayers by simple exfoliation from their bulk counterparts highly probable.  The ferromagnetic nature of in-plane Cu-Cu exchanges, driven by the antiferrodistortive arrangement of the Jahn Teller active Cu ions in the square coordination of Cl ions, remains intact on exfoliation of 2D monolayers from bulk crystals. Their calculated single-ion anisotropy shows it to be an order of magnitude higher than that of CrGeTe$_3$ which is an established 2D ferromagnet\cite{Gong2017}, thereby explaining the possible stabilisation of ferromagnetism in 2D by breakdown of Mermin Wagner theorem \cite{Mermin66}. They found the anisotropy for the hybrid Cu perovskite systems to be of an easy-plane type rather than an easy-axis type as found for CrGeTe$_3$. The finite-temperature ordering of such an easy-plane, spin-1/2 Cu spin was shown to be achievable within the formulation of a spin Hamiltonian with a generalized form of anisotropy, with the off-diagonal terms of the anisotropy matrix being responsible for driving the ordering at finite temperature.
 
Another interesting route of inducing 2D magnetism in hybrid organic inorganic perovskites, particularly in the $p$ bands of the $X^-$ radicals may be by hole doping or electrostatic doping by application of moderate electric fields which dopes the Fermi level of strongly hybridised $p$ levels and due to Stoner instability gives rise to half metallic 2D ferromagnetism in $p$ orbitals of $X^-$ ions. This has been shown in the context of 2D magnetism in general \cite{Banerjee_2021_2D}, however may prove to be a strong route of tunability in 2D hybrid perovskites and may be explored both theoretically and experimentally.

\subsection{\textcolor{black}{{Light Dependent Transitions in Metal-Organic Complexes}}}
In this section we discuss the light dependent studies carried out on SCO materials, in particular LIESST (Light-Induced Excited Spin-State Trapping) which happens to be a method of changing the electronic spin state of a compound by means of irradiation with light \cite{Ohkoshi2011, balde, hauser, anna, ramasesha, daubric, Marshak2012, hauser2, Liu2013, Bertoni2015}. Due to the majority of SCO complexes being based on iron we also focus in this section on the iron based complexes. Experimental studies have been primarily based on such iron based compounds.
For iron complexes, LIESST involves excitation of the low spin complex with green light to a triplet state. Two successive steps of inter-system crossing result in the high spin complex. Movement from the high spin complex to the low spin complex requires excitation with red light. We shall briefly discuss the various studies in detail in this section.

The study of LIESST started with a legendary article by P G\"{u}tlich and A. Hauser \cite{GUTLICH19901}, in the 1990s, which demonstrated that at very low temperatures, the low spin state ($^1A_1$) in a Fe(II) complex can be converted quantitatively to the high spin state ($^5T_2$)  by irradiating the sample into the $^1A_1 \rightarrow ^1T_1$ d-d absorption band ($\sim$540 nm). The HS state obtained, although meta stable, however has a very long lifetime at low temperatures, and did not show any noticeable decay over a period of several days at 10 K. Thermal relaxation back to the LS state occurred only beyond a critical temperature. Reconversion to the LS state was possible by irradiating into the $^5T_2 \rightarrow ^5E$ absorption band ($\sim$850 nm). Hence the system was seen to behave like an optical switch. The mechanism of the transition was well described. Lasers can induce a singlet state transition which produce short-lived states. However the alternate mechanism due to the presence of spin orbit coupling in the system can induce a singlet to triplet state transition. As long as the energy difference between the HS and LS potential surfaces, which are well separated by the large difference in metal-ligand bond length, cannot be overcome thermally the system remains trapped in the HS state with a very long lifetime. 

Indeed several studies both theoretical and computational highlight this particular mechanism of the light induced transition being able to follow several pathways between different excited spin states. It has been shown that just as metal-organic complexes show pressure driven spin state transition as demonstrated before, they can be made to undergo light induced spin state trapping \cite{Karmakar2022}. A computational study of the light-induced excited spin-state trapping (LIESST) in a number of Fe(II) spin
crossover complexes, coordinated by monodentate, bidentate and multidentate ligands was carried out, with the goal to uncover the trend in the low temperature relaxation rate. A very larger order of magnitude change in low temperature relaxation rate was observed among the complexes \citenum{Karmakar2022}. The basic mechanism of light induced spin state trapping requires the presence of presence of spin states with difference potential energy wells situated at different energy scales and presence of more than one excitation pathway as shown in Fig. 7. This particular idea works for the metal-organic complexes very well. Coupled with the phenomena of light absorption and storage if they can be made to undergo a spin state switching a huge variety of application possibilities will open up instantly for this particular field of study. One may show very effectively here that energy saving considerations shall be particularly important. If light induced switching can control devices which store energy obtained from solar energy this may be the next giant step in the evolution of energy saving technologies

\begin{figure}[h!]
	\centering
	\includegraphics[width=0.5\columnwidth]{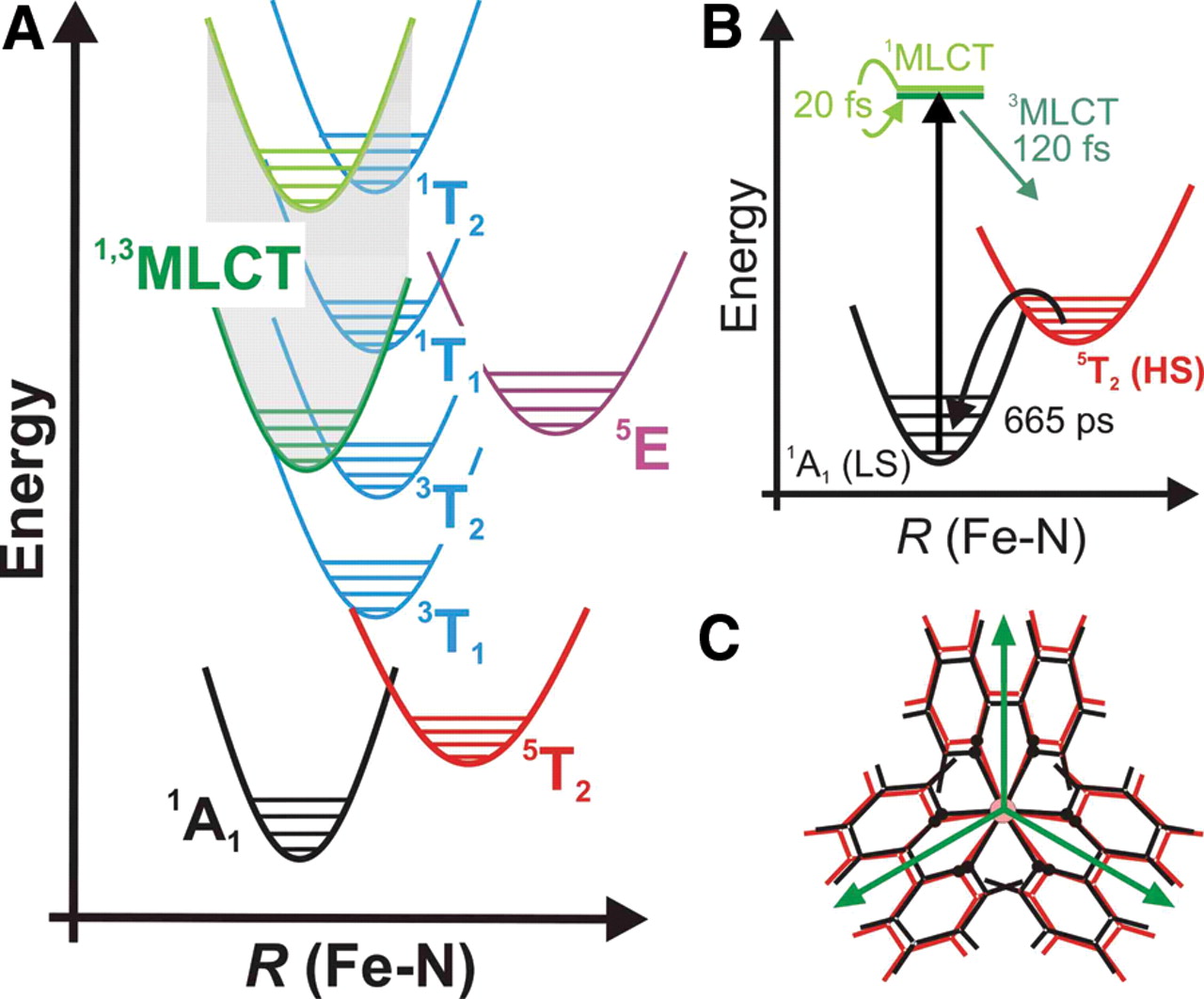}
	\caption{Figure showing mechanism of LIESST in Fe based metal organic complex. Figure adopted from Bressler et al.., \cite{Bressler489}}
\end{figure}

\subsection{\textcolor{black}{{Light Dependent Transitions in Hybrid Perovskites}}}

\begin{figure}[h!]
	\begin{center}
		\includegraphics[width=\columnwidth]{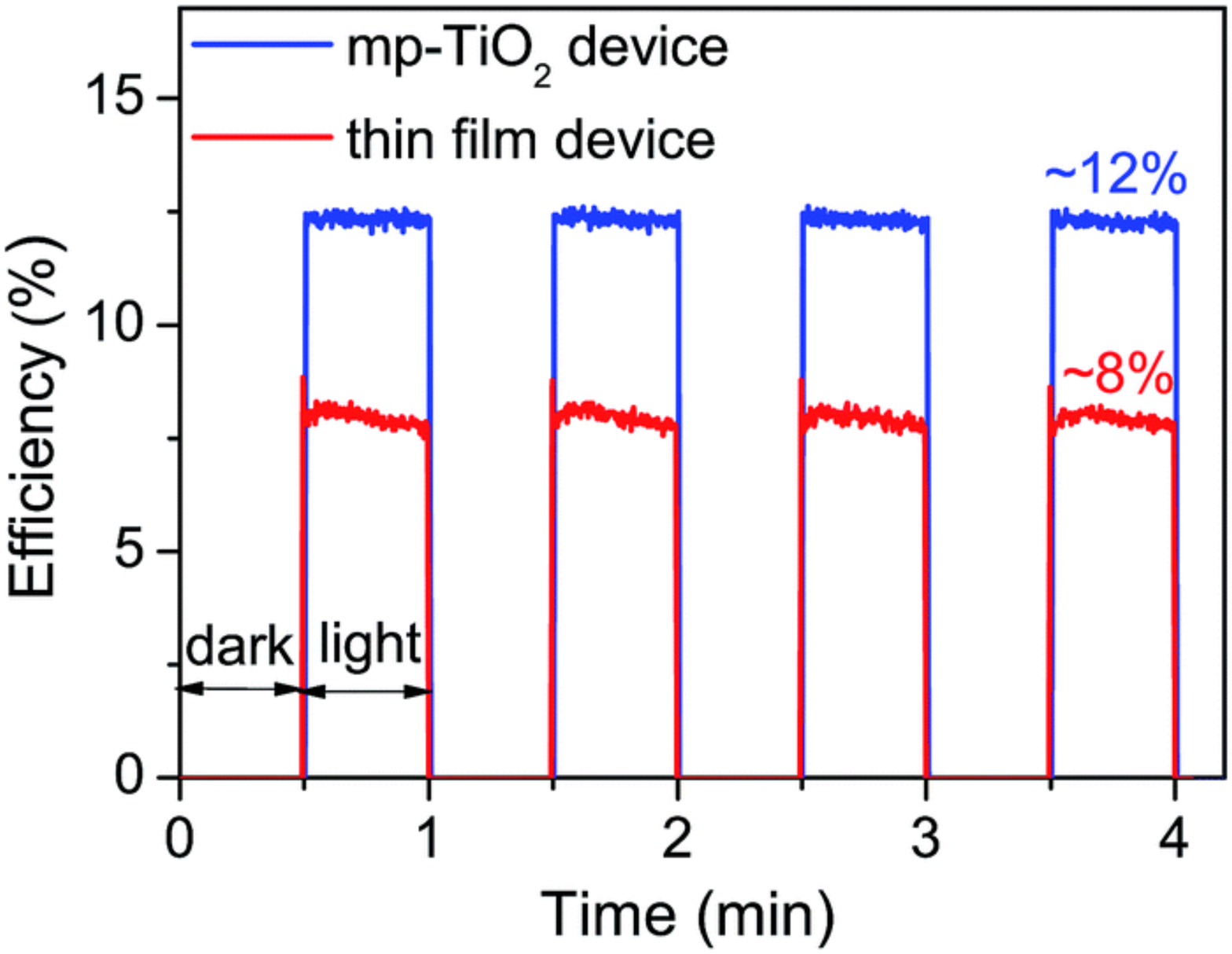}
		\caption{Effect of light illumination on the conversion efficiency of devices with different architectures. Reprinted with permission from ref \citenum{Unger}. Copyright 2016 American Institute of Physics.  }
	\end{center}
	\label{Light}
\end{figure}

\textcolor{black}{The halide based hybrid perovskites show a high propensity towards light induced transitions simply due to the experimental observations of their usage in solar cells. It is in context of these particular studies that one can proceed further in reference to energy saving and harvesting explorations. Hybrid perovskites have shown an immense performance boost in the recent years in the field of solar cell applications. If combined with the potential applications in the field of light induced spin crossover particularly spin state switching with the potentials in energy storage in the form of solar cells, LEDs with the switching capabilities, hybrid perovskites can be one of the biggest discoveries in terms of energy efficient multifarious application based materials, which is the need of the hour. We have already discussed hybrid perovskites in context of solar cells and also how they can be shown to undergo spin state transition at least theoretically. In this section we first deal with the general theory of light induced spin crossover and finally try to give an idea as to why hybrid perovskites can also be ideal candidates in this case. To the best of our knowledge, no study exists in the field yet which has correlated hybrid perovskites with light induced spin transition.}

\textcolor{black}{Measuring the efficiency of the perovskite solar cell after light illumination is called light soaking. In this section, we will see how light soaking has a positive impact on the perovskite systems. Illuminating light on the perovskite solar cells seems to increase the efficiency of the system. Docampo et al. reported that upon light illumination of \ce{CH3NH3PbI_{3-x}Cl_x} the PCE\cite{Docampo2013} was observed to be little over $>1$\% PCE from the current-voltage curve. But after operation in air and illumination of light for 10 minutes, the PCE increased to 7.5-10\%. But for regular MSSC, the PCE increases immediately. Light soaking is also done for \ce{TiO_x}\cite{Docampo2013}. \ce{CH3NH3PbI_{3-x}Cl_x} based perovskite solar cells\cite{Barrows} were fabricated using ultra-sonic spray coating. During the first 20 minutes of light illumination, PCE was enhanced implying a positive light soaking effect. The solar cells were at open-circuit while illumination. It led to an increase in device FF, Jsc, and Voc. Due to the presence of \ce{TiO_x}, a light soaking effect is observed. But as there was an absence of \ce{TiO_x} in this device, the light soaking phenomena may be due to mechanistically different processes\cite{Barrows}.}

\textcolor{black}{Unger et al. compared solution processed thin film devices to meso-porous titania based perovskite devices\cite{Unger}. When the external stimulus of light was illuminated on the device with periodically switching the stimulus on and off, it was observed that the thin film gave PCE of 8\%(Figure \ref{Light}). On constant illumination, light on the thin films showed a slow decrease of photocurrent. From the IV measurements, it was reported that the PCE of meso-porous titania based perovskite was 12\%. The hysteretic effects during IV-measurements showed that light had a major role to play in the transient processes. The thin films showed low performance when IV curves were measured after storage in the dark before illuminating with light. The PCE increased from 2\% to 4\% after light-soaking of the device at far-forward bias conditions(J$>$0). It was due to the elevation of fill factor and photocurrent. However for mesoporous titania based devices, there was not much difference in the efficiency before and after light illumination\cite{Unger}.}

\textcolor{black}{Liu et al. synthesised planar and mesoscopic perovskite thin film by vapour-assisted solution process (VASP)\cite{Liu}. It was absorbed by the structure before and after light illumination remained the same as seen from the XRD pattern. Light soaking in the forward bias increases FF and Voc as interposed in the previous reports. This lead to an increase in PCE by 1.5 times. However in  case of light soaking in reverse bias, a decrease in the FF and Voc was observed. It was due to the fact that under reverse bias situation the flow of photocurrent was suppressed by remnant polarisation, but for the forward bias situation, the electric field and remnant polarisation field were in the same direction, because of which the photocurrents separated and transported more efficiently\cite{Liu}. Zhao et al. observed the effect of light as an external stimulus on indium tin oxide (ITO)/ poly(3,4-ethylene dioxythiophene) poly(styrene-sulfonate) (PEDOT:PSS)/  \ce{CH3NH3PbI_{3-x}Cl_x}/[6,6]-phenyl C61-butyric acid methyl ester\\ (PCBM)/aluminum (Al) devices \cite{Zhao}. Reversible light soaking phenomena were observed in the perovskite solar cells. It was reported that, on the illumination with light, the open-circuit voltage(Voc) and the fill-factor(FF) increases. On exposure of light, short-circuit current(Jsc) initially increases but again decreases. The charge accumulation at the electrode interfaces decreased on light soaking as observed from the C-V measurements. This phenomenon can be due to neutralizing of the inter-facial defects at the electrode interface due to photo-generated carrier upon light illumination. Also, the built-in electric field changed due to the migration of ions which in turn affected the charge accumulation at the electrode interface. There was a decrease in the bulk electrical polarisation by light soaking as observed by frequency dependent capacitance\cite{Zhao}. It is worth mentioning about \ce{(FAPbI3)0.85(MA-PbBr3)0.15} perovskite\cite{Kim2019}, which showed structural changes under illumination. The experimental investigations identified the corresponding nanoscale ferroelastic domains. At the phase transition temperature, it was observed that the domains disappeared and reappeared by temperature dependent topography measurements. The domain patterns were visible in both amplitude and phase under the light stimulus. There was also an increase in the structural disorder under light illumination\cite{Kim2019}.
}

\subsection{Spin Transition Induced Optical Properties in Metal-Organic Complexes and Hybrid Perovskites}
\begin{figure}
	\centering
	\includegraphics[width=0.6\columnwidth]{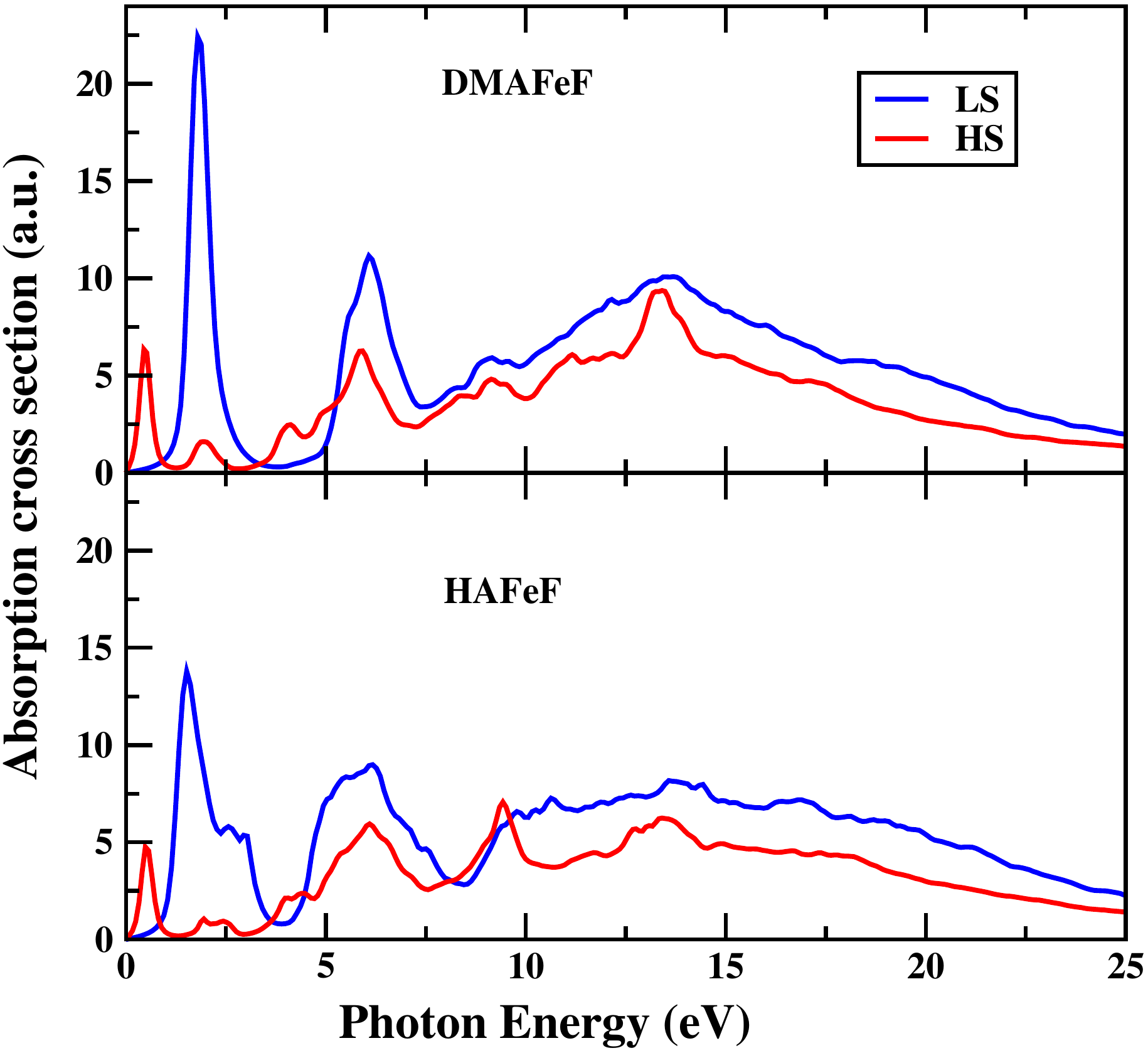}
	\caption{Figure showing absorption cross section with photon energies. The HS and LS states show different absorption cross section and they have distinctive features for two different materials HAFeF and DMAFeF.}
	\label{opt-spect}
\end{figure}
In this section we indicate that one may explore several routes to optical properties design induced by the spin crossover phenomena. In figure \ref{opt-spect} we show the absorption cross section for DMAFeF and HAFeF, for the two different HS and LS spin states. The absorption cross section for both the materials are very different in HS and LS states. This leads to the phenomena of piezochromism in such materials as described before as well since the HS and LS states can be switched by using both pressure and temperature. This also indicates that the materials have different colour spectra in different spin states since their absorption spectra are different for the spin states. Another possible implication is that the spin states may also be switched with appropriate absorption of photon energies, which (without getting into the complex time dependent DFT calculations) is indicative of a light induced spin state switching in hybrid perovskites. Piezochromism has also been shown for hybrid perovskites experimentally but has not been associated with any spin crossover phenomena. From our study we indicate that both piezochromism associated with spin crossover and LIESST are possible in case of hybrid perovskites and may be explored in much more detail.

\textcolor{black}{It is to be noted here that the idea of piezochromism being associated with spin transitions is relatively new. In a very recent work \cite{banerjee2021} Banerjee et al have shown the possibility of piezochromism in a Mn based hybrid perovskite being associated with spin transitions and magnetic hysteresis. Although this idea is relatively new, there have been instances of colour change of materials being associated with spin transitions in spin crossover coordination polymers like Fe triazole. Now since such spin transitions albeit mostly shown to be experimentally driven by temperature, may also in principle be driven by application of pressure as well. Hence a direct correlation may be drawn between piezochromism and spin transitions exists. We demonstrate this here for two hybrid perovskites DMAFeF and HAFeF, which has not been seen before.}
\begin{figure}[h!]
	\begin{center}
		\includegraphics[width=0.6\columnwidth]{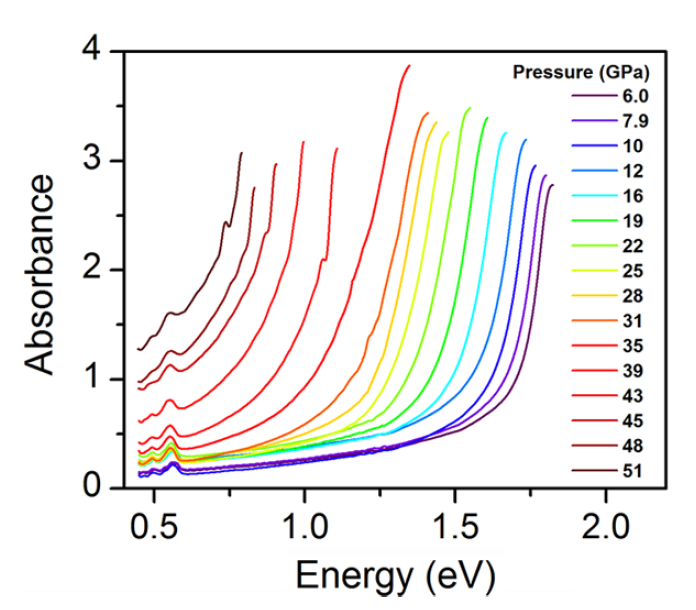}
		\caption{Experimental observation of Piezochromism in Hybrid Perovskites. Figure adapted from Jaffe et al.. \cite{Jaffe2015}}
	\end{center}
\end{figure}
External pressure driven changes in the structural and electronic properties open the field of tuning materials properties through compression. High pressure in the scale of gigapascals (GPa) affects a wide range of structural, electronic, magnetic and optical transformation in such materials. As mentioned previously , piezochromism [As shown in the Figure 10] is coming out as another exciting phenomena, where the optical properties change dramatically with the external pressure, in such metal organic complexes. For example, the high-pressure phases of Cu-Cl perovskites \cite{Jaffe2015} reveal the octahedral rotations and bond compression in the inorganic layers under the hydrostatic pressure and this can lead to the piezochromism with varying electronic conductivity. The lattice compression certainly promote the exotic behaviour like enhanced superconductivity, magnetic ordering, and metalisation in such hybrid perovskites.  The high pressure research also provide important experimental calibration for theoretical models of electronic properties based on the inter-atomic distances, which can be systematically varied with lattice compression. 
\textcolor{black}{It is worth mentioning here that although Cu-Cl does indeed show piezochromism it is not a candidate for Spin crossover in $d^9$ or $d^10$ configuration. In case of spin crossover materials one is primarily interested in 3$d$ transition metals in $d^4$ to $d^7$ configuration. It has recently been shown in a theoretical study that formate based hybrid perovskites which can show spin transition are also candidates which can show piezochromism under the influence of pressure. \cite{banerjee2021} It is also to be mentioned that although there have been several suggestions regarding spin transitions in formate based perovskites, there are hardly any examples of halide based perovskites showing spin transitions. This in part may primarily be due to the fact that most halide based perovskites are Cu, Sn, or Pb based perovskites which are not ideal candidates for spin crossover owing to their electronic configurations.}

\section{\textcolor{black}{{Summary and Future Directions}}}
External stimuli driven transitions in hybrid perovskites and metal organic complexes need more attention both from the perspective of fundamental understanding of the materials properties as well as improving the performances of the optoelectronic devices based on them.  Spin orbit interaction is an important parameter that governs the electronic structure and subsequently the magnetic and optical properties of such perovskite materials \cite{Chakraborty2021, Kaur2022}. This should be addressed while performing computation despite the fact that they are computationally expensive. As most of the hybrid perovskite structures are highly sensitive and decompose in presence of moisture or air, hence the corresponding stability is a serious factor as far the lifetime of the device is concerned. The optical band gap is another sensitive issue as far the simulation is concerned in such hybrid perovskites. As band gap and optical response is an excited state property, therefore only ground state calculations with simplified exchange correlation functional may not give the accurate bad gap value and as a consequence the desired optical response. Incorporating the derivative discontinuity of the Kohn-Sham orbital during the transition from ground state to excited state modeling, should be considered as hybrid exchange correlation functional. As the correction factor is mainly hidden within the exchange correlation functional, therefore the choice of the accurate and reliable functional is always required in order to have a better understanding of the optical properties of such hybrid perovskites. Much deeper chemical insight from defect formation review is required. 

The spin-dependent band structure of hybrid perovskite paves the way to manipulate the spin that leads to the potential applications in the field of spintronics and spin-orbitronics. Computational results  give an improved description of the band structures as a consequence one can also predict Rashba or Dresselhaus spin splitting or both in these hybrid systems \cite{Kepenekian2015, Kepenekian2017, Chakraborty2021}. Spin crossover in hybrid perovskites may occur upon light irradiation or by application of magnetic field, as observed in many of the metal-organic complexes. {\textcolor{black}{A wide range of applications starting from spin-orbitronics to energy scavenging in the form of solar cells and light emitting diodes (LEDs) can be envisaged through the tuning of Rashba-Dresselhaus effect, which can be further  influenced through the external stimuli driven transitions by pressure, temperature and light in hybrid perovskites and metal organic complexes.}} Light induced spin-switching, if can be achieved in hybrid perovskites, will make these materials useful in application as optical switches as well as for studying phenomena as light-induced excited spin-state trapping. It is expected that this study will stimulate further research, both experimental and theoretical, in discovering novel hybrid perovskites.For such versatile area of research as mentioned above, a profound understanding of the fundamental properties like mechanical, optical, magnetic, ferroelectric is much needed and required a substantial scientific effort to explore extensively. The atomistic insight of a material is quite relevant in order to explore its inherent different properties and thus electronic structure calculations become more essential day by day. 
\\

\section{Acknowledgement}
The authors would like to acknowledge HRI Allahabad for the infrastructure and SRG/2020/001707 for funding.
\\

\textcolor{black}{H.B. and J.K. have contributed equally to this work.}


\bibliography{biblio-paper.bib}

\begin{thebibliography}{100}
\expandafter\ifx\csname url\endcsname\relax
  \def\url#1{\texttt{#1}}\fi
\expandafter\ifx\csname urlprefix\endcsname\relax\def\urlprefix{URL }\fi
\expandafter\ifx\csname href\endcsname\relax
  \def\href#1#2{#2} \def\path#1{#1}\fi

\bibitem{metal-org1}
H.~Xu, R.~Chen, Q.~Sun, W.~Lai, Q.~Su, W.~Huang, X.~Liu,
  \href{http://dx.doi.org/10.1039/C3CS60449G}{Recent progress in
  metal–organic complexes for optoelectronic applications}, Chem. Soc. Rev.
  43 (2014) 3259--3302.
\newblock \href {https://doi.org/10.1039/C3CS60449G}
  {\path{doi:10.1039/C3CS60449G}}.
\newline\urlprefix\url{http://dx.doi.org/10.1039/C3CS60449G}

\bibitem{metal-org2}
E.~L. Shock, C.~M. Koretsky,
  \href{http://www.sciencedirect.com/science/article/pii/0016703795000588}{Metal-organic
  complexes in geochemical processes: Estimation of standard partial molal
  thermodynamic properties of aqueous complexes between metal cations and
  monovalent organic acid ligands at high pressures and temperatures},
  Geochimica et Cosmochimica Acta 59~(8) (1995) 1497 -- 1532.
\newblock \href {https://doi.org/https://doi.org/10.1016/0016-7037(95)00058-8}
  {\path{doi:https://doi.org/10.1016/0016-7037(95)00058-8}}.
\newline\urlprefix\url{http://www.sciencedirect.com/science/article/pii/0016703795000588}

\bibitem{stuart}
S.~R. Batten, N.~R. Champness, X.-M. Chen, J.~Garcia-Martinez, S.~Kitagawa,
  L.~Öhrström, M.~O’Keeffe, M.~P. Suh, J.~Reedijk,
  \href{https://www.degruyter.com/view/journals/pac/85/8/article-p1715.xml}{Terminology
  of metal–organic frameworks and coordination polymers (iupac
  recommendations 2013)}, Pure and Applied Chemistry 85~(8) (2013) 1715 --
  1724.
\newblock \href {https://doi.org/https://doi.org/10.1351/PAC-REC-12-11-20}
  {\path{doi:https://doi.org/10.1351/PAC-REC-12-11-20}}.
\newline\urlprefix\url{https://www.degruyter.com/view/journals/pac/85/8/article-p1715.xml}

\bibitem{cote}
A.~P. C{\^o}t{\'e}, A.~I. Benin, N.~W. Ockwig, M.~O{\textquoteright}Keeffe,
  A.~J. Matzger, O.~M. Yaghi,
  \href{https://science.sciencemag.org/content/310/5751/1166}{Porous,
  crystalline, covalent organic frameworks}, Science 310~(5751) (2005)
  1166--1170.
\newblock \href
  {http://arxiv.org/abs/https://science.sciencemag.org/content/310/5751/1166.full.pdf}
  {\path{arXiv:https://science.sciencemag.org/content/310/5751/1166.full.pdf}},
  \href {https://doi.org/10.1126/science.1120411}
  {\path{doi:10.1126/science.1120411}}.
\newline\urlprefix\url{https://science.sciencemag.org/content/310/5751/1166}

\bibitem{sensors}
J.~Linares, E.~Codjovi, Y.~Garcia,
  \href{https://www.mdpi.com/1424-8220/12/4/4479}{Pressure and temperature spin
  crossover sensors with optical detection}, Sensors 12~(4) (2012) 4479--4492.
\newblock \href {https://doi.org/10.3390/s120404479}
  {\path{doi:10.3390/s120404479}}.
\newline\urlprefix\url{https://www.mdpi.com/1424-8220/12/4/4479}

\bibitem{catalysis}
A.~U. Czaja, N.~Trukhan, U.~Müller,
  \href{http://dx.doi.org/10.1039/B804680H}{Industrial applications of
  metal–organic frameworks}, Chem. Soc. Rev. 38 (2009) 1284--1293.
\newblock \href {https://doi.org/10.1039/B804680H}
  {\path{doi:10.1039/B804680H}}.
\newline\urlprefix\url{http://dx.doi.org/10.1039/B804680H}

\bibitem{biomedical}
P.~Horcajada, T.~Chalati, C.~Serre, B.~Gillet, C.~Sebrie, T.~Baati, J.~F.
  Eubank, D.~Heurtaux, P.~Clayette, C.~Kreuz, J.-S. Chang, Y.~K. Hwang,
  V.~Marsaud, P.-N. Bories, L.~Cynober, S.~Gil, G.~F{\'e}rey, P.~Couvreur,
  R.~Gref, \href{https://doi.org/10.1038/nmat2608}{Porous
  metal-organic-framework nanoscale carriers as a potential platform for drug
  delivery?and imaging}, Nature Materials 9~(2) (2010) 172--178.
\newblock \href {https://doi.org/10.1038/nmat2608}
  {\path{doi:10.1038/nmat2608}}.
\newline\urlprefix\url{https://doi.org/10.1038/nmat2608}

\bibitem{morris}
R.~E. Morris, P.~S. Wheatley,
  \href{http://dx.doi.org/10.1002/anie.200703934}{Gas storage in nanoporous
  materials}, Angewandte Chemie International Edition 47~(27) (2008)
  4966--4981.
\newblock \href {https://doi.org/10.1002/anie.200703934}
  {\path{doi:10.1002/anie.200703934}}.
\newline\urlprefix\url{http://dx.doi.org/10.1002/anie.200703934}

\bibitem{kreno}
L.~E. Kreno, K.~Leong, O.~K. Farha, M.~Allendorf, R.~P. Van~Duyne, J.~T. Hupp,
  Metal–organic framework materials as chemical sensors, Chemical Reviews
  112~(2) (2012) 1105--1125.
\newblock \href {https://doi.org/10.1021/cr200324t}
  {\path{doi:10.1021/cr200324t}}.

\bibitem{absorp}
H.~Xu, R.~Chen, Q.~Sun, W.~Lai, Q.~Su, W.~Huang, X.~Liu,
  \href{http://dx.doi.org/10.1039/C3CS60449G}{Recent progress in
  metal–organic complexes for optoelectronic applications}, Chem. Soc. Rev.
  43 (2014) 3259--3302.
\newblock \href {https://doi.org/10.1039/C3CS60449G}
  {\path{doi:10.1039/C3CS60449G}}.
\newline\urlprefix\url{http://dx.doi.org/10.1039/C3CS60449G}

\bibitem{mobility1}
T.~C. Narayan, T.~Miyakai, S.~Seki, M.~Dincă,
  \href{https://doi.org/10.1021/ja3059827}{High charge mobility in a
  tetrathiafulvalene-based microporous metal–organic framework}, Journal of
  the American Chemical Society 134~(31) (2012) 12932--12935, pMID: 22827709.
\newblock \href {http://arxiv.org/abs/https://doi.org/10.1021/ja3059827}
  {\path{arXiv:https://doi.org/10.1021/ja3059827}}, \href
  {https://doi.org/10.1021/ja3059827} {\path{doi:10.1021/ja3059827}}.
\newline\urlprefix\url{https://doi.org/10.1021/ja3059827}

\bibitem{mobility2}
C.~Yang, R.~Dong, M.~Wang, P.~S. Petkov, Z.~Zhang, M.~Wang, P.~Han,
  M.~Ballabio, S.~A. Br{\"a}uninger, Z.~Liao, J.~Zhang, F.~Schwotzer,
  E.~Zschech, H.-H. Klauss, E.~C{\'a}novas, S.~Kaskel, M.~Bonn, S.~Zhou,
  T.~Heine, X.~Feng, \href{https://doi.org/10.1038/s41467-019-11267-w}{A
  semiconducting layered metal-organic framework magnet}, Nature Communications
  10~(1) (2019) 3260.
\newblock \href {https://doi.org/10.1038/s41467-019-11267-w}
  {\path{doi:10.1038/s41467-019-11267-w}}.
\newline\urlprefix\url{https://doi.org/10.1038/s41467-019-11267-w}

\bibitem{excitonbe}
M.~Baranowski, P.~Plochocka,
  \href{https://onlinelibrary.wiley.com/doi/abs/10.1002/aenm.201903659}{Excitons
  in metal-halide perovskites}, Advanced Energy Materials 10~(26) (2020)
  1903659.
\newblock \href
  {http://arxiv.org/abs/https://onlinelibrary.wiley.com/doi/pdf/10.1002/aenm.201903659}
  {\path{arXiv:https://onlinelibrary.wiley.com/doi/pdf/10.1002/aenm.201903659}},
  \href {https://doi.org/10.1002/aenm.201903659}
  {\path{doi:10.1002/aenm.201903659}}.
\newline\urlprefix\url{https://onlinelibrary.wiley.com/doi/abs/10.1002/aenm.201903659}

\bibitem{excitondl}
B.~T. Luppi, D.~Majak, M.~Gupta, E.~Rivard, K.~Shankar,
  \href{http://dx.doi.org/10.1039/C8TA10037C}{Triplet excitons: improving
  exciton diffusion length for enhanced organic photovoltaics}, J. Mater. Chem.
  A 7 (2019) 2445--2463.
\newblock \href {https://doi.org/10.1039/C8TA10037C}
  {\path{doi:10.1039/C8TA10037C}}.
\newline\urlprefix\url{http://dx.doi.org/10.1039/C8TA10037C}

\bibitem{Cheetham}
G.~Kieslich, S.~Sun, A.~K. Cheetham,
  \href{http://dx.doi.org/10.1039/C4SC02211D}{Solid-state principles applied to
  organic–inorganic perovskites: new tricks for an old dog}, Chem. Sci. 5
  (2014) 4712--4715.
\newblock \href {https://doi.org/10.1039/C4SC02211D}
  {\path{doi:10.1039/C4SC02211D}}.
\newline\urlprefix\url{http://dx.doi.org/10.1039/C4SC02211D}

\bibitem{hybrid-formate}
Z.~Wang, K.~Hu, S.~Gao, H.~Kobayashi,
  \href{https://onlinelibrary.wiley.com/doi/abs/10.1002/adma.200904438}{Formate-based
  magnetic metal–organic frameworks templated by protonated amines}, Advanced
  Materials 22~(13) (2010) 1526--1533.
\newblock \href
  {http://arxiv.org/abs/https://onlinelibrary.wiley.com/doi/pdf/10.1002/adma.200904438}
  {\path{arXiv:https://onlinelibrary.wiley.com/doi/pdf/10.1002/adma.200904438}},
  \href {https://doi.org/10.1002/adma.200904438}
  {\path{doi:10.1002/adma.200904438}}.
\newline\urlprefix\url{https://onlinelibrary.wiley.com/doi/abs/10.1002/adma.200904438}

\bibitem{Clune2020}
A.~Clune, N.~Harms, K.~R. O'Neal, K.~Hughey, K.~A. Smith, D.~Obeysekera,
  J.~Haddock, N.~S. Dalal, J.~Yang, Z.~Liu, J.~L. Musfeldt,
  \href{https://doi.org/10.1021/acs.inorgchem.0c01225}{Developing the
  pressure-temperature-magnetic field phase diagram of multiferroic
  [(ch3)2nh2]mn(hcoo)3}, Inorganic Chemistry 59~(14) (2020) 10083--10090.
\newblock \href {https://doi.org/10.1021/acs.inorgchem.0c01225}
  {\path{doi:10.1021/acs.inorgchem.0c01225}}.
\newline\urlprefix\url{https://doi.org/10.1021/acs.inorgchem.0c01225}

\bibitem{Sanchez-Andujar2010}
M.~Sanchez-Andujar, S.~Presedo, S.~Yanez-Vilar, S.~Castro-Garcia, J.~Shamir,
  M.~A. Senaris-Rodriguez,
  \href{https://doi.org/10.1021/ic901872g}{Characterization of the
  order-disorder dielectric transition in the hybrid organic-inorganic
  perovskite-like formate mn(hcoo)3[(ch3)2nh2]}, Inorganic Chemistry 49~(4)
  (2010) 1510--1516.
\newblock \href {https://doi.org/10.1021/ic901872g}
  {\path{doi:10.1021/ic901872g}}.
\newline\urlprefix\url{https://doi.org/10.1021/ic901872g}

\bibitem{Chitnis2018}
A.~V. Chitnis, H.~Bhatt, M.~Maczka, M.~N. Deo, N.~Garg,
  \href{http://dx.doi.org/10.1039/C8DT03080D}{Remarkable resilience of the
  formate cage in a multiferroic metal organic framework material: dimethyl
  ammonium manganese formate (dmamnf)}, Dalton Trans. 47 (2018) 12993--13005.
\newblock \href {https://doi.org/10.1039/C8DT03080D}
  {\path{doi:10.1039/C8DT03080D}}.
\newline\urlprefix\url{http://dx.doi.org/10.1039/C8DT03080D}

\bibitem{letard}
J.-F. Letard, L.~Capes, G.~Chastanet, N.~Moliner, S.~Letard, J.-A. Real,
  O.~Kahn,
  \href{http://www.sciencedirect.com/science/article/pii/S0009261499010362}{Critical
  temperature of the liesst effect in iron(ii) spin crossover compounds},
  Chemical Physics Letters 313~(1) (1999) 115 -- 120.
\newblock \href
  {https://doi.org/http://dx.doi.org/10.1016/S0009-2614(99)01036-2}
  {\path{doi:http://dx.doi.org/10.1016/S0009-2614(99)01036-2}}.
\newline\urlprefix\url{http://www.sciencedirect.com/science/article/pii/S0009261499010362}

\bibitem{Letard2004}
J.-F. L{\'e}tard, P.~Guionneau, L.~Goux-Capes,
  \href{https://doi.org/10.1007/b95429}{Towards Spin Crossover Applications},
  Springer Berlin Heidelberg, Berlin, Heidelberg, 2004, pp. 221--249.
\newblock \href {https://doi.org/10.1007/b95429} {\path{doi:10.1007/b95429}}.
\newline\urlprefix\url{https://doi.org/10.1007/b95429}

\bibitem{cobo}
S.~Cobo, G.~Molnar, J.~A. Real, A.~Bousseksou,
  \href{http://dx.doi.org/10.1002/anie.200601885}{Multilayer sequential
  assembly of thin films that display room-temperature spin crossover with
  hysteresis}, Angewandte Chemie International Edition 45~(35) (2006)
  5786--5789.
\newblock \href {https://doi.org/10.1002/anie.200601885}
  {\path{doi:10.1002/anie.200601885}}.
\newline\urlprefix\url{http://dx.doi.org/10.1002/anie.200601885}

\bibitem{brooker}
S.~Brooker, J.~A. Kitchen,
  \href{http://dx.doi.org/10.1039/B907682D}{Nano-magnetic materials: spin
  crossover compounds vs. single molecule magnets vs. single chain magnets},
  Dalton Trans. (2009) 7331--7340\href {https://doi.org/10.1039/B907682D}
  {\path{doi:10.1039/B907682D}}.
\newline\urlprefix\url{http://dx.doi.org/10.1039/B907682D}

\bibitem{halder}
G.~J. Halder, C.~J. Kepert, B.~Moubaraki, K.~S. Murray, J.~D. Cashion,
  \href{http://science.sciencemag.org/content/298/5599/1762}{Guest-dependent
  spin crossover in a nanoporous molecular framework material}, Science
  298~(5599) (2002) 1762--1765.
\newblock \href
  {http://arxiv.org/abs/http://science.sciencemag.org/content/298/5599/1762.full.pdf}
  {\path{arXiv:http://science.sciencemag.org/content/298/5599/1762.full.pdf}},
  \href {https://doi.org/10.1126/science.1075948}
  {\path{doi:10.1126/science.1075948}}.
\newline\urlprefix\url{http://science.sciencemag.org/content/298/5599/1762}

\bibitem{Jain1}
P.~Jain, N.~S. Dalal, B.~H. Toby, H.~W. Kroto, A.~K. Cheetham,
  \href{https://doi.org/10.1021/ja801952e}{Order-disorder antiferroelectric
  phase transition in a hybrid inorganic-organic framework with the perovskite
  architecture}, Journal of the American Chemical Society 130~(32) (2008)
  10450--10451.
\newblock \href {https://doi.org/10.1021/ja801952e}
  {\path{doi:10.1021/ja801952e}}.
\newline\urlprefix\url{https://doi.org/10.1021/ja801952e}

\bibitem{Jain2}
P.~Jain, V.~Ramachandran, R.~J. Clark, H.~D. Zhou, B.~H. Toby, N.~S. Dalal,
  H.~W. Kroto, A.~K. Cheetham,
  \href{https://doi.org/10.1021/ja904156s}{Multiferroic behavior associated
  with an order-disorder hydrogen bonding transition in metal-organic
  frameworks (mofs) with the perovskite abx3 architecture}, Journal of the
  American Chemical Society 131~(38) (2009) 13625--13627.
\newblock \href {https://doi.org/10.1021/ja904156s}
  {\path{doi:10.1021/ja904156s}}.
\newline\urlprefix\url{https://doi.org/10.1021/ja904156s}

\bibitem{stroppa1}
A.~Stroppa, P.~Jain, P.~Barone, M.~Marsman, J.~M. Perez-Mato, A.~K. Cheetham,
  H.~W. Kroto, S.~Picozzi,
  \href{http://dx.doi.org/10.1002/anie.201101405}{Electric control of
  magnetization and interplay between orbital ordering and ferroelectricity in
  a multiferroic metal–organic framework}, Angewandte Chemie International
  Edition 50~(26) (2011) 5847--5850.
\newblock \href {https://doi.org/10.1002/anie.201101405}
  {\path{doi:10.1002/anie.201101405}}.
\newline\urlprefix\url{http://dx.doi.org/10.1002/anie.201101405}

\bibitem{stroppa2}
A.~Stroppa, P.~Barone, P.~Jain, J.~M. Perez-Mato, S.~Picozzi,
  \href{http://dx.doi.org/10.1002/adma.201204738}{Hybrid improper
  ferroelectricity in a multiferroic and magnetoelectric metal-organic
  framework}, Advanced Materials 25~(16) (2013) 2284--2290.
\newblock \href {https://doi.org/10.1002/adma.201204738}
  {\path{doi:10.1002/adma.201204738}}.
\newline\urlprefix\url{http://dx.doi.org/10.1002/adma.201204738}

\bibitem{stroppa4}
Y.~Tian, A.~Stroppa, Y.-S. Chai, P.~Barone, M.~Perez-Mato, S.~Picozzi, Y.~Sun,
  \href{http://dx.doi.org/10.1002/pssr.201409470}{High-temperature
  ferroelectricity and strong magnetoelectric effects in a hybrid
  organic–inorganic perovskite framework}, physica status solidi (RRL) -
  Rapid Research Letters 9~(1) (2015) 62--67.
\newblock \href {https://doi.org/10.1002/pssr.201409470}
  {\path{doi:10.1002/pssr.201409470}}.
\newline\urlprefix\url{http://dx.doi.org/10.1002/pssr.201409470}

\bibitem{stroppa5}
W.-P. Zhao, C.~Shi, A.~Stroppa, D.~Di~Sante, F.~Cimpoesu, W.~Zhang,
  \href{http://dx.doi.org/10.1021/acs.inorgchem.6b01545}{Lone-pair-electron-driven
  ionic displacements in a ferroelectric metal–organic hybrid}, Inorganic
  Chemistry 55~(20) (2016) 10337--10342, pMID: 27676140.
\newblock \href
  {http://arxiv.org/abs/http://dx.doi.org/10.1021/acs.inorgchem.6b01545}
  {\path{arXiv:http://dx.doi.org/10.1021/acs.inorgchem.6b01545}}, \href
  {https://doi.org/10.1021/acs.inorgchem.6b01545}
  {\path{doi:10.1021/acs.inorgchem.6b01545}}.
\newline\urlprefix\url{http://dx.doi.org/10.1021/acs.inorgchem.6b01545}

\bibitem{Jain2016}
P.~Jain, A.~Stroppa, D.~Nabok, A.~Marino, A.~Rubano, D.~Paparo, M.~Matsubara,
  H.~Nakotte, M.~Fiebig, S.~Picozzi, E.~S. Choi, A.~K. Cheetham, C.~Draxl,
  N.~S. Dalal, V.~S. Zapf,
  \href{http://dx.doi.org/10.1038/npjquantmats.2016.12}{Switchable electric
  polarization and ferroelectric domains in a metal-organic-framework} 1 (2016)
  16012 EP --, article.
\newline\urlprefix\url{http://dx.doi.org/10.1038/npjquantmats.2016.12}

\bibitem{sante}
D.~Di~Sante, A.~Stroppa, P.~Jain, S.~Picozzi,
  \href{http://dx.doi.org/10.1021/ja408283a}{Tuning the ferroelectric
  polarization in a multiferroic metal–organic framework}, Journal of the
  American Chemical Society 135~(48) (2013) 18126--18130, pMID: 24191632.
\newblock \href {http://arxiv.org/abs/http://dx.doi.org/10.1021/ja408283a}
  {\path{arXiv:http://dx.doi.org/10.1021/ja408283a}}, \href
  {https://doi.org/10.1021/ja408283a} {\path{doi:10.1021/ja408283a}}.
\newline\urlprefix\url{http://dx.doi.org/10.1021/ja408283a}

\bibitem{Tian2014}
Y.~Tian, A.~Stroppa, Y.~Chai, L.~Yan, S.~Wang, P.~Barone, S.~Picozzi, Y.~Sun,
  \href{http://dx.doi.org/10.1038/srep06062}{Cross coupling between electric
  and magnetic orders in a multiferroic metal-organic framework} 4 (2014) 6062
  EP --, article.
\newline\urlprefix\url{http://dx.doi.org/10.1038/srep06062}

\bibitem{aguirre}
L.~C. Gómez-Aguirre, B.~Pato-Doldán, A.~Stroppa, S.~Yáñez-Vilar,
  L.~Bayarjargal, B.~Winkler, S.~Castro-García, J.~Mira, M.~Sánchez-Andújar,
  M.~A. Señarís-Rodríguez,
  \href{http://dx.doi.org/10.1021/ic502218n}{Room-temperature polar order in
  [nh4][cd(hcoo)3] - a hybrid inorganic–organic compound with a unique
  perovskite architecture}, Inorganic Chemistry 54~(5) (2015) 2109--2116, pMID:
  25664382.
\newblock \href {http://arxiv.org/abs/http://dx.doi.org/10.1021/ic502218n}
  {\path{arXiv:http://dx.doi.org/10.1021/ic502218n}}, \href
  {https://doi.org/10.1021/ic502218n} {\path{doi:10.1021/ic502218n}}.
\newline\urlprefix\url{http://dx.doi.org/10.1021/ic502218n}

\bibitem{kamminga}
M.~E. Kamminga, A.~Stroppa, S.~Picozzi, M.~Chislov, I.~A. Zvereva, J.~Baas,
  A.~Meetsma, G.~R. Blake, T.~T.~M. Palstra,
  \href{http://dx.doi.org/10.1021/acs.inorgchem.6b01699}{Polar nature of
  (ch3nh3)3bi2i9 perovskite-like hybrids}, Inorganic Chemistry 56~(1) (2017)
  33--41, pMID: 27626290.
\newblock \href
  {http://arxiv.org/abs/http://dx.doi.org/10.1021/acs.inorgchem.6b01699}
  {\path{arXiv:http://dx.doi.org/10.1021/acs.inorgchem.6b01699}}, \href
  {https://doi.org/10.1021/acs.inorgchem.6b01699}
  {\path{doi:10.1021/acs.inorgchem.6b01699}}.
\newline\urlprefix\url{http://dx.doi.org/10.1021/acs.inorgchem.6b01699}

\bibitem{ptak}
M.~Ptak, M.~Maczka, A.~Gagor, A.~Sieradzki, A.~Stroppa, D.~Di~Sante, J.~M.
  Perez-Mato, L.~Macalik,
  \href{http://dx.doi.org/10.1039/C5DT04536C}{Experimental and theoretical
  studies of structural phase transition in a novel polar perovskite-like
  [c2h5nh3][na0.5fe0.5(hcoo)3] formate}, Dalton Trans. 45 (2016) 2574--2583.
\newblock \href {https://doi.org/10.1039/C5DT04536C}
  {\path{doi:10.1039/C5DT04536C}}.
\newline\urlprefix\url{http://dx.doi.org/10.1039/C5DT04536C}

\bibitem{ghosh}
S.~Ghosh, D.~Di~Sante, A.~Stroppa,
  \href{http://dx.doi.org/10.1021/acs.jpclett.5b01806}{Strain tuning of
  ferroelectric polarization in hybrid organic inorganic perovskite compounds},
  The Journal of Physical Chemistry Letters 6~(22) (2015) 4553--4559, pMID:
  26512946.
\newblock \href
  {http://arxiv.org/abs/http://dx.doi.org/10.1021/acs.jpclett.5b01806}
  {\path{arXiv:http://dx.doi.org/10.1021/acs.jpclett.5b01806}}, \href
  {https://doi.org/10.1021/acs.jpclett.5b01806}
  {\path{doi:10.1021/acs.jpclett.5b01806}}.
\newline\urlprefix\url{http://dx.doi.org/10.1021/acs.jpclett.5b01806}

\bibitem{mazzuca}
L.~Mazzuca, L.~Cañadillas-Delgado, J.~A. Rodríguez-Velamazán, O.~Fabelo,
  M.~Scarrozza, A.~Stroppa, S.~Picozzi, J.-P. Zhao, X.-H. Bu,
  J.~Rodríguez-Carvajal,
  \href{http://dx.doi.org/10.1021/acs.inorgchem.6b01866}{Magnetic structures of
  heterometallic m(ii)–m(iii) formate compounds}, Inorganic Chemistry 56~(1)
  (2017) 197--207, pMID: 27935298.
\newblock \href
  {http://arxiv.org/abs/http://dx.doi.org/10.1021/acs.inorgchem.6b01866}
  {\path{arXiv:http://dx.doi.org/10.1021/acs.inorgchem.6b01866}}, \href
  {https://doi.org/10.1021/acs.inorgchem.6b01866}
  {\path{doi:10.1021/acs.inorgchem.6b01866}}.
\newline\urlprefix\url{http://dx.doi.org/10.1021/acs.inorgchem.6b01866}

\bibitem{Banerjee2014}
H.~Banerjee, M.~Kumar, T.~Saha-Dasgupta,
  \href{https://link.aps.org/doi/10.1103/PhysRevB.90.174433}{Cooperativity in
  spin-crossover transition in metalorganic complexes: Interplay of magnetic
  and elastic interactions}, Phys. Rev. B 90 (2014) 174433.
\newblock \href {https://doi.org/10.1103/PhysRevB.90.174433}
  {\path{doi:10.1103/PhysRevB.90.174433}}.
\newline\urlprefix\url{https://link.aps.org/doi/10.1103/PhysRevB.90.174433}

\bibitem{Banerjee2016}
H.~Banerjee, S.~Chakraborty, T.~Saha-Dasgupta,
  \href{https://doi.org/10.1021/acs.chemmater.6b03755}{Cationic effect on
  pressure driven spin-state transition and cooperativity in hybrid
  perovskites}, Chemistry of Materials 28~(22) (2016) 8379--8384.
\newblock \href {https://doi.org/10.1021/acs.chemmater.6b03755}
  {\path{doi:10.1021/acs.chemmater.6b03755}}.
\newline\urlprefix\url{https://doi.org/10.1021/acs.chemmater.6b03755}

\bibitem{Banerjee2017}
H.~Banerjee, S.~Chakraborty, T.~Saha-Dasgupta,
  \href{https://www.mdpi.com/2304-6740/5/3/47}{Design and control of
  cooperativity in spin-crossover in metal–organic complexes: A theoretical
  overview}, Inorganics 5~(3) (2017).
\newblock \href {https://doi.org/10.3390/inorganics5030047}
  {\path{doi:10.3390/inorganics5030047}}.
\newline\urlprefix\url{https://www.mdpi.com/2304-6740/5/3/47}

\bibitem{banerjee2021}
H.~Banerjee, A.~Rittsteuer, M.~Aichhorn,
  \href{https://link.aps.org/doi/10.1103/PhysRevMaterials.6.044401}{Temperature
  and pressure driven spin transitions and piezochromism in a mn-based hybrid
  perovskite}, Phys. Rev. Materials 6 (2022) 044401.
\newblock \href {https://doi.org/10.1103/PhysRevMaterials.6.044401}
  {\path{doi:10.1103/PhysRevMaterials.6.044401}}.
\newline\urlprefix\url{https://link.aps.org/doi/10.1103/PhysRevMaterials.6.044401}

\bibitem{Chakraborty2017}
S.~Chakraborty, W.~Xie, N.~Mathews, M.~Sherburne, R.~Ahuja, M.~Asta, S.~G.
  Mhaisalkar, \href{https://doi.org/10.1021/acsenergylett.7b00035}{Rational
  design: A high-throughput computational screening and experimental validation
  methodology for lead-free and emergent hybrid perovskites}, ACS Energy
  Letters 2~(4) (2017) 837--845.
\newblock \href {https://doi.org/10.1021/acsenergylett.7b00035}
  {\path{doi:10.1021/acsenergylett.7b00035}}.
\newline\urlprefix\url{https://doi.org/10.1021/acsenergylett.7b00035}

\bibitem{Delgado2012}
L.~Canadillas-Delgado, O.~Fabelo, J.~A. Rodriguez-Velamazan, M.-H.
  Lemee-Cailleau, S.~A. Mason, E.~Pardo, F.~Lloret, J.-P. Zhao, X.-H. Bu,
  V.~Simonet, C.~V. Colin, J.~Rodriguez-Carvajal,
  \href{https://doi.org/10.1021/ja3082457}{The role of order-disorder
  transitions in the quest for molecular multiferroics: Structural and magnetic
  neutron studies of a mixed valence iron(ii)-iron(iii) formate framework},
  Journal of the American Chemical Society 134~(48) (2012) 19772--19781.
\newblock \href {https://doi.org/10.1021/ja3082457}
  {\path{doi:10.1021/ja3082457}}.
\newline\urlprefix\url{https://doi.org/10.1021/ja3082457}

\bibitem{sudipto-sco1}
K.~Tarafder, S.~Kanungo, P.~M. Oppeneer, T.~Saha-Dasgupta,
  \href{https://link.aps.org/doi/10.1103/PhysRevLett.109.077203}{Pressure and
  temperature control of spin-switchable metal-organic coordination polymers
  from ab initio calculations}, Phys. Rev. Lett. 109 (2012) 077203.
\newblock \href {https://doi.org/10.1103/PhysRevLett.109.077203}
  {\path{doi:10.1103/PhysRevLett.109.077203}}.
\newline\urlprefix\url{https://link.aps.org/doi/10.1103/PhysRevLett.109.077203}

\bibitem{sudipto-sco2}
P.~Maldonado, S.~Kanungo, T.~Saha-Dasgupta, P.~M. Oppeneer,
  \href{https://link.aps.org/doi/10.1103/PhysRevB.88.020408}{Two-step
  spin-switchable tetranuclear fe(ii) molecular solid: Ab initio theory and
  predictions}, Phys. Rev. B 88 (2013) 020408.
\newblock \href {https://doi.org/10.1103/PhysRevB.88.020408}
  {\path{doi:10.1103/PhysRevB.88.020408}}.
\newline\urlprefix\url{https://link.aps.org/doi/10.1103/PhysRevB.88.020408}

\bibitem{tsd-rev}
T.~Saha-Dasgupta, P.~M. Oppeneer, Computational design of magnetic
  metal-organic complexes and coordination polymers with spin-switchable
  functionalities, MRS Bulletin 39~(7) (2014) 614--620.
\newblock \href {https://doi.org/10.1557/mrs.2014.112}
  {\path{doi:10.1557/mrs.2014.112}}.

\bibitem{carignano2014thermal}
M.~A. Carignano, A.~Kachmar, J.~Hutter, Thermal effects on ch$_3$nh$_3$pbi$_3$
  perovskite from ab-initio molecular dynamics simulations (2014).
\newblock \href {http://arxiv.org/abs/1409.6842} {\path{arXiv:1409.6842}}.

\bibitem{Shivam}
S.~Singh, C.~Li, F.~Panzer, K.~L. Narasimhan, A.~Graeser, T.~P. Gujar,
  A.~Köhler, M.~Thelakkat, S.~Huettner, D.~Kabra,
  \href{https://doi.org/10.1021/acs.jpclett.6b01207}{Effect of thermal and
  structural disorder on the electronic structure of hybrid perovskite
  semiconductor ch3nh3pbi3}, The Journal of Physical Chemistry Letters 7~(15)
  (2016) 3014--3021, pMID: 27435936.
\newblock \href
  {http://arxiv.org/abs/https://doi.org/10.1021/acs.jpclett.6b01207}
  {\path{arXiv:https://doi.org/10.1021/acs.jpclett.6b01207}}, \href
  {https://doi.org/10.1021/acs.jpclett.6b01207}
  {\path{doi:10.1021/acs.jpclett.6b01207}}.
\newline\urlprefix\url{https://doi.org/10.1021/acs.jpclett.6b01207}

\bibitem{Olgui2002}
D.~Olguin, M.~Cardona, A.~Cantarero, Electron-phonon effects on the direct band
  gap in semiconductors: Lcao calculations, Solid State Communications 122~(11)
  (2002) 575--589.

\bibitem{Varshni1967TemperatureDO}
Y.~P. Varshni, Temperature dependence of the energy gap in semiconductors,
  Physica D: Nonlinear Phenomena 34 (1967) 149--154.

\bibitem{Innocenzo2014}
V.~D'Innocenzo, G.~Grancini, M.~J.~P. Alcocer, A.~R.~S. Kandada, S.~D. Stranks,
  M.~M. Lee, G.~Lanzani, H.~J. Snaith, A.~Petrozza,
  \href{https://doi.org/10.1038/ncomms4586}{Excitons versus free charges in
  organo-lead tri-halide perovskites}, Nature Communications 5~(1) (2014) 3586.
\newblock \href {https://doi.org/10.1038/ncomms4586}
  {\path{doi:10.1038/ncomms4586}}.
\newline\urlprefix\url{https://doi.org/10.1038/ncomms4586}

\bibitem{Docampo2013}
P.~Docampo, J.~M. Ball, M.~Darwich, G.~E. Eperon, H.~J. Snaith,
  \href{https://doi.org/10.1038/ncomms3761}{Efficient organometal trihalide
  perovskite planar-heterojunction solar cells on flexible polymer substrates},
  Nature Communications 4~(1) (2013) 2761.
\newblock \href {https://doi.org/10.1038/ncomms3761}
  {\path{doi:10.1038/ncomms3761}}.
\newline\urlprefix\url{https://doi.org/10.1038/ncomms3761}

\bibitem{Park}
X.~Zhao, H.-S. Kim, J.-Y. Seo, N.-G. Park,
  \href{https://doi.org/10.1021/acsami.6b15673}{Effect of selective contacts on
  the thermal stability of perovskite solar cells}, ACS Applied Materials \&
  Interfaces 9~(8) (2017) 7148--7153, pMID: 28186718.
\newblock \href {http://arxiv.org/abs/https://doi.org/10.1021/acsami.6b15673}
  {\path{arXiv:https://doi.org/10.1021/acsami.6b15673}}, \href
  {https://doi.org/10.1021/acsami.6b15673} {\path{doi:10.1021/acsami.6b15673}}.
\newline\urlprefix\url{https://doi.org/10.1021/acsami.6b15673}

\bibitem{Liyuan}
Y.~Wu, F.~Xie, H.~Chen, X.~Yang, H.~Su, M.~Cai, Z.~Zhou, T.~Noda, L.~Han,
  \href{https://onlinelibrary.wiley.com/doi/abs/10.1002/adma.201701073}{Thermally
  stable mapbi3 perovskite solar cells with efficiency of 19.19\% and area over
  1 cm2 achieved by additive engineering}, Advanced Materials 29~(28) (2017)
  1701073.
\newblock \href
  {http://arxiv.org/abs/https://onlinelibrary.wiley.com/doi/pdf/10.1002/adma.201701073}
  {\path{arXiv:https://onlinelibrary.wiley.com/doi/pdf/10.1002/adma.201701073}},
  \href {https://doi.org/https://doi.org/10.1002/adma.201701073}
  {\path{doi:https://doi.org/10.1002/adma.201701073}}.
\newline\urlprefix\url{https://onlinelibrary.wiley.com/doi/abs/10.1002/adma.201701073}

\bibitem{Meng2021}
Q.~Meng, Y.~Chen, Y.~Y. Xiao, J.~Sun, X.~Zhang, C.~B. Han, H.~Gao, Y.~Zhang,
  H.~Yan, \href{https://doi.org/10.1007/s10854-020-03029-y}{Effect of
  temperature on the performance of perovskite solar cells}, Journal of
  Materials Science: Materials in Electronics 32~(10) (2021) 12784--12792.
\newblock \href {https://doi.org/10.1007/s10854-020-03029-y}
  {\path{doi:10.1007/s10854-020-03029-y}}.
\newline\urlprefix\url{https://doi.org/10.1007/s10854-020-03029-y}

\bibitem{Bressler489}
C.~Bressler, C.~Milne, V.-T. Pham, A.~ElNahhas, R.~M. van~der Veen, W.~Gawelda,
  S.~Johnson, P.~Beaud, D.~Grolimund, M.~Kaiser, C.~N. Borca, G.~Ingold,
  R.~Abela, M.~Chergui,
  \href{https://science.sciencemag.org/content/323/5913/489}{Femtosecond xanes
  study of the light-induced spin crossover dynamics in an iron(ii) complex},
  Science 323~(5913) (2009) 489--492.
\newblock \href
  {http://arxiv.org/abs/https://science.sciencemag.org/content/323/5913/489.full.pdf}
  {\path{arXiv:https://science.sciencemag.org/content/323/5913/489.full.pdf}},
  \href {https://doi.org/10.1126/science.1165733}
  {\path{doi:10.1126/science.1165733}}.
\newline\urlprefix\url{https://science.sciencemag.org/content/323/5913/489}

\bibitem{Adam}
A.~Jaffe, Y.~Lin, W.~L. Mao, H.~I. Karunadasa,
  \href{https://doi.org/10.1021/ja512396m}{Pressure-induced conductivity and
  yellow-to-black piezochromism in a layered cu–cl hybrid perovskite},
  Journal of the American Chemical Society 137~(4) (2015) 1673--1678, pMID:
  25580620.
\newblock \href {http://arxiv.org/abs/https://doi.org/10.1021/ja512396m}
  {\path{arXiv:https://doi.org/10.1021/ja512396m}}, \href
  {https://doi.org/10.1021/ja512396m} {\path{doi:10.1021/ja512396m}}.
\newline\urlprefix\url{https://doi.org/10.1021/ja512396m}

\bibitem{Capitani}
F.~Capitani, C.~Marini, S.~Caramazza, P.~Postorino, G.~Garbarino, M.~Hanfland,
  A.~Pisanu, P.~Quadrelli, L.~Malavasi,
  \href{https://doi.org/10.1063/1.4948577}{High-pressure behavior of
  methylammonium lead iodide (mapbi3) hybrid perovskite}, Journal of Applied
  Physics 119~(18) (2016) 185901.
\newblock \href {http://arxiv.org/abs/https://doi.org/10.1063/1.4948577}
  {\path{arXiv:https://doi.org/10.1063/1.4948577}}, \href
  {https://doi.org/10.1063/1.4948577} {\path{doi:10.1063/1.4948577}}.
\newline\urlprefix\url{https://doi.org/10.1063/1.4948577}

\bibitem{Jaffe2015}
A.~Jaffe, Y.~Lin, W.~L. Mao, H.~I. Karunadasa,
  \href{https://doi.org/10.1021/ja512396m}{Pressure-induced conductivity and
  yellow-to-black piezochromism in a layered cu-cl hybrid perovskite}, Journal
  of the American Chemical Society 137~(4) (2015) 1673--1678.
\newblock \href {https://doi.org/10.1021/ja512396m}
  {\path{doi:10.1021/ja512396m}}.
\newline\urlprefix\url{https://doi.org/10.1021/ja512396m}

\bibitem{Umeyama2016}
D.~Umeyama, Y.~Lin, H.~I. Karunadasa,
  \href{https://doi.org/10.1021/acs.chemmater.6b01147}{Red-to-black
  piezochromism in a compressible pb-i-scn layered perovskite}, Chemistry of
  Materials 28~(10) (2016) 3241--3244.
\newblock \href {https://doi.org/10.1021/acs.chemmater.6b01147}
  {\path{doi:10.1021/acs.chemmater.6b01147}}.
\newline\urlprefix\url{https://doi.org/10.1021/acs.chemmater.6b01147}

\bibitem{Kojima2009}
A.~Kojima, K.~Teshima, Y.~Shirai, T.~Miyasaka,
  \href{https://doi.org/10.1021/ja809598r}{Organometal halide perovskites as
  visible-light sensitizers for photovoltaic cells}, Journal of the American
  Chemical Society 131~(17) (2009) 6050--6051.
\newblock \href {https://doi.org/10.1021/ja809598r}
  {\path{doi:10.1021/ja809598r}}.
\newline\urlprefix\url{https://doi.org/10.1021/ja809598r}

\bibitem{Giovanni2015}
D.~Giovanni, H.~Ma, J.~Chua, M.~Gratzel, R.~Ramesh, S.~Mhaisalkar, N.~Mathews,
  T.~C. Sum, \href{https://doi.org/10.1021/nl5039314}{Highly spin-polarized
  carrier dynamics and ultralarge photoinduced magnetization in ch3nh3pbi3
  perovskite thin films}, Nano Letters 15~(3) (2015) 1553--1558.
\newblock \href {https://doi.org/10.1021/nl5039314}
  {\path{doi:10.1021/nl5039314}}.
\newline\urlprefix\url{https://doi.org/10.1021/nl5039314}

\bibitem{Savory2016}
C.~N. Savory, A.~Walsh, D.~O. Scanlon,
  \href{https://doi.org/10.1021/acsenergylett.6b00471}{Can pb-free halide
  double perovskites support high-efficiency solar cells?}, ACS Energy Letters
  1~(5) (2016) 949--955.
\newblock \href {https://doi.org/10.1021/acsenergylett.6b00471}
  {\path{doi:10.1021/acsenergylett.6b00471}}.
\newline\urlprefix\url{https://doi.org/10.1021/acsenergylett.6b00471}

\bibitem{Swainson2007}
I.~P. Swainson, M.~G. Tucker, D.~J. Wilson, B.~Winkler, V.~Milman,
  \href{https://doi.org/10.1021/cm0621601}{Pressure response of an
  organic-inorganic perovskite: Methylammonium lead bromide}, Chemistry of
  Materials 19~(10) (2007) 2401--2405.
\newblock \href {https://doi.org/10.1021/cm0621601}
  {\path{doi:10.1021/cm0621601}}.
\newline\urlprefix\url{https://doi.org/10.1021/cm0621601}

\bibitem{arnab-piezo}
A.~Majumdar, A.~A. Adeleke, S.~Chakraborty, R.~Ahuja,
  \href{http://dx.doi.org/10.1039/D0TC04516K}{Emerging piezochromism in lead
  free alkaline earth chalcogenide perovskite azrs3 (a = mg{,} ca{,} sr and ba)
  under pressure}, J. Mater. Chem. C 8 (2020) 16392--16403.
\newblock \href {https://doi.org/10.1039/D0TC04516K}
  {\path{doi:10.1039/D0TC04516K}}.
\newline\urlprefix\url{http://dx.doi.org/10.1039/D0TC04516K}

\bibitem{arnab-piezo-2}
A.~Majumdar, S.~Chakraborty, R.~Ahuja,
  \href{https://doi.org/10.1063/5.0019132}{Emerging piezochromism in
  transparent lead free perovskite rb3x2i9 (x=sb, bi) under compression: A
  comparative theoretical insight}, Journal of Applied Physics 128~(4) (2020)
  045102.
\newblock \href {http://arxiv.org/abs/https://doi.org/10.1063/5.0019132}
  {\path{arXiv:https://doi.org/10.1063/5.0019132}}, \href
  {https://doi.org/10.1063/5.0019132} {\path{doi:10.1063/5.0019132}}.
\newline\urlprefix\url{https://doi.org/10.1063/5.0019132}

\bibitem{TCO}
J.~H. Heo, D.~S. Lee, D.~H. Shin, S.~H. Im,
  \href{http://dx.doi.org/10.1039/C8TA09452G}{Recent advancements in and
  perspectives on flexible hybrid perovskite solar cells}, J. Mater. Chem. A 7
  (2019) 888--900.
\newblock \href {https://doi.org/10.1039/C8TA09452G}
  {\path{doi:10.1039/C8TA09452G}}.
\newline\urlprefix\url{http://dx.doi.org/10.1039/C8TA09452G}

\bibitem{Lee}
Y.~Lee, D.~B. Mitzi, P.~W. Barnes, T.~Vogt,
  \href{https://link.aps.org/doi/10.1103/PhysRevB.68.020103}{Pressure-induced
  phase transitions and templating effect in three-dimensional
  organic-inorganic hybrid perovskites}, Phys. Rev. B 68 (2003) 020103.
\newblock \href {https://doi.org/10.1103/PhysRevB.68.020103}
  {\path{doi:10.1103/PhysRevB.68.020103}}.
\newline\urlprefix\url{https://link.aps.org/doi/10.1103/PhysRevB.68.020103}

\bibitem{Kenji}
K.~Ohwada, K.~Ishii, T.~Inami, Y.~Murakami, T.~Shobu, H.~Ohsumi, N.~Ikeda,
  Y.~Ohishi,
  \href{https://link.aps.org/doi/10.1103/PhysRevB.72.014123}{Structural
  properties and phase transition of hole-orbital-ordered
  ${({\mathrm{C}}_{2}{\mathrm{H}}_{5}\mathrm{N}{\mathrm{H}}_{3})}_{2}\mathrm{Cu}{\mathrm{cl}}_{4}$
  studied by resonant and non-resonant x-ray scatterings under high pressure},
  Phys. Rev. B 72 (2005) 014123.
\newblock \href {https://doi.org/10.1103/PhysRevB.72.014123}
  {\path{doi:10.1103/PhysRevB.72.014123}}.
\newline\urlprefix\url{https://link.aps.org/doi/10.1103/PhysRevB.72.014123}

\bibitem{Swainson}
I.~P. Swainson, M.~G. Tucker, D.~J. Wilson, B.~Winkler, V.~Milman, Pressure
  response of an organic-inorganic perovskite: Methylammonium lead bromide,
  Chemistry of Materials 19~(10) (2007) 2401--2405.
\newblock \href {https://doi.org/10.1021/cm0621601}
  {\path{doi:10.1021/cm0621601}}.

\bibitem{Wang}
Y.~Wang, X.~Lü, W.~Yang, T.~Wen, L.~Yang, X.~Ren, L.~Wang, Z.~Lin, Y.~Zhao,
  \href{https://doi.org/10.1021/jacs.5b06346}{Pressure-induced phase
  transformation, reversible amorphization, and anomalous visible light
  response in organolead bromide perovskite}, Journal of the American Chemical
  Society 137~(34) (2015) 11144--11149, pMID: 26284441.
\newblock \href {http://arxiv.org/abs/https://doi.org/10.1021/jacs.5b06346}
  {\path{arXiv:https://doi.org/10.1021/jacs.5b06346}}, \href
  {https://doi.org/10.1021/jacs.5b06346} {\path{doi:10.1021/jacs.5b06346}}.
\newline\urlprefix\url{https://doi.org/10.1021/jacs.5b06346}

\bibitem{Tianji}
T.~Ou, J.~Yan, C.~Xiao, W.~Shen, C.~Liu, X.~Liu, Y.~Han, Y.~Ma, C.~Gao,
  \href{http://dx.doi.org/10.1039/C5NR07842C}{Visible light response{,}
  electrical transport{,} and amorphization in compressed organolead iodine
  perovskites}, Nanoscale 8 (2016) 11426--11431.
\newblock \href {https://doi.org/10.1039/C5NR07842C}
  {\path{doi:10.1039/C5NR07842C}}.
\newline\urlprefix\url{http://dx.doi.org/10.1039/C5NR07842C}

\bibitem{wang2015pressureinduced}
K.~Wang, R.~Liu, Y.~Qiao, J.~Cui, B.~Song, B.~Liu, B.~Zou, Pressure-induced
  reversible phase transition and amorphization of ch$_3$nh$_3$pbi$_3$ (2015).
\newblock \href {http://arxiv.org/abs/1509.03717} {\path{arXiv:1509.03717}}.

\bibitem{Zou}
L.~Wang, K.~Wang, B.~Zou,
  \href{https://doi.org/10.1021/acs.jpclett.6b00999}{Pressure-induced
  structural and optical properties of organometal halide perovskite-based
  formamidinium lead bromide}, The Journal of Physical Chemistry Letters 7~(13)
  (2016) 2556--2562, pMID: 27321024.
\newblock \href
  {http://arxiv.org/abs/https://doi.org/10.1021/acs.jpclett.6b00999}
  {\path{arXiv:https://doi.org/10.1021/acs.jpclett.6b00999}}, \href
  {https://doi.org/10.1021/acs.jpclett.6b00999}
  {\path{doi:10.1021/acs.jpclett.6b00999}}.
\newline\urlprefix\url{https://doi.org/10.1021/acs.jpclett.6b00999}

\bibitem{Yurong}
X.~Wang, H.~Tian, X.~Li, H.~Sang, C.~Zhong, J.-M. Liu, Y.~Yang,
  \href{http://dx.doi.org/10.1039/D0CP05892K}{Pressure effects on the
  structures and electronic properties of halide perovskite cspbx3 (x = i{,}
  br{,} cl)}, Phys. Chem. Chem. Phys. 23 (2021) 3479--3484.
\newblock \href {https://doi.org/10.1039/D0CP05892K}
  {\path{doi:10.1039/D0CP05892K}}.
\newline\urlprefix\url{http://dx.doi.org/10.1039/D0CP05892K}

\bibitem{Chakraborty2021}
S.~Chakraborty, M.~K. Nazeeruddin,
  \href{https://doi.org/10.1021/acs.jpclett.0c02497}{The status quo of rashba
  phenomena in organic--inorganic hybrid perovskites}, The Journal of Physical
  Chemistry Letters 12~(1) (2021) 361--367.
\newblock \href {https://doi.org/10.1021/acs.jpclett.0c02497}
  {\path{doi:10.1021/acs.jpclett.0c02497}}.
\newline\urlprefix\url{https://doi.org/10.1021/acs.jpclett.0c02497}

\bibitem{Kaur2022}
J.~Kaur, S.~Chakraborty, \href{https://doi.org/10.1021/acsaem.1c03824}{Tuning
  spin texture and spectroscopic limited maximum efficiency through chemical
  composition space in double halide perovskites}, ACS Applied Energy Materials
  (Apr 2022).
\newblock \href {https://doi.org/10.1021/acsaem.1c03824}
  {\path{doi:10.1021/acsaem.1c03824}}.
\newline\urlprefix\url{https://doi.org/10.1021/acsaem.1c03824}

\bibitem{Guloy}
D.~B. Mitzi, S.~Wang, C.~A. Feild, C.~A. Chess, A.~M. Guloy,
  \href{https://www.science.org/doi/abs/10.1126/science.267.5203.1473}{Conducting
  layered organic-inorganic halides containing \&\#x3008;110\&\#x3009;-oriented
  perovskite sheets}, Science 267~(5203) (1995) 1473--1476.
\newblock \href
  {http://arxiv.org/abs/https://www.science.org/doi/pdf/10.1126/science.267.5203.1473}
  {\path{arXiv:https://www.science.org/doi/pdf/10.1126/science.267.5203.1473}},
  \href {https://doi.org/10.1126/science.267.5203.1473}
  {\path{doi:10.1126/science.267.5203.1473}}.
\newline\urlprefix\url{https://www.science.org/doi/abs/10.1126/science.267.5203.1473}

\bibitem{Dohner}
E.~R. Dohner, E.~T. Hoke, H.~I. Karunadasa,
  \href{https://doi.org/10.1021/ja411045r}{Self-assembly of broadband
  white-light emitters}, Journal of the American Chemical Society 136~(5)
  (2014) 1718--1721, pMID: 24422494.
\newblock \href {http://arxiv.org/abs/https://doi.org/10.1021/ja411045r}
  {\path{arXiv:https://doi.org/10.1021/ja411045r}}, \href
  {https://doi.org/10.1021/ja411045r} {\path{doi:10.1021/ja411045r}}.
\newline\urlprefix\url{https://doi.org/10.1021/ja411045r}

\bibitem{Ballouli}
A.~O. El-Ballouli, O.~M. Bakr, O.~F. Mohammed,
  \href{https://doi.org/10.1021/acs.jpclett.0c00359}{Structurally tunable
  two-dimensional layered perovskites: From confinement and enhanced charge
  transport to prolonged hot carrier cooling dynamics}, The Journal of Physical
  Chemistry Letters 11~(14) (2020) 5705--5718, pMID: 32574063.
\newblock \href
  {http://arxiv.org/abs/https://doi.org/10.1021/acs.jpclett.0c00359}
  {\path{arXiv:https://doi.org/10.1021/acs.jpclett.0c00359}}, \href
  {https://doi.org/10.1021/acs.jpclett.0c00359}
  {\path{doi:10.1021/acs.jpclett.0c00359}}.
\newline\urlprefix\url{https://doi.org/10.1021/acs.jpclett.0c00359}

\bibitem{Xujie}
M.~Li, T.~Liu, Y.~Wang, W.~Yang, X.~Lü,
  \href{https://doi.org/10.1063/1.5133653}{Pressure responses of halide
  perovskites with various compositions, dimensionalities, and morphologies},
  Matter and Radiation at Extremes 5~(1) (2020) 018201.
\newblock \href {http://arxiv.org/abs/https://doi.org/10.1063/1.5133653}
  {\path{arXiv:https://doi.org/10.1063/1.5133653}}, \href
  {https://doi.org/10.1063/1.5133653} {\path{doi:10.1063/1.5133653}}.
\newline\urlprefix\url{https://doi.org/10.1063/1.5133653}

\bibitem{Manasa}
M.~G. Basavarajappa, M.~K. Nazeeruddin, S.~Chakraborty,
  \href{https://doi.org/10.1063/5.0053128}{Evolution of hybrid
  organic–inorganic perovskite materials under external pressure}, Applied
  Physics Reviews 8~(4) (2021) 041309.
\newblock \href {http://arxiv.org/abs/https://doi.org/10.1063/5.0053128}
  {\path{arXiv:https://doi.org/10.1063/5.0053128}}, \href
  {https://doi.org/10.1063/5.0053128} {\path{doi:10.1063/5.0053128}}.
\newline\urlprefix\url{https://doi.org/10.1063/5.0053128}

\bibitem{Mohanty_2022}
P.~P. Mohanty, R.~Ahuja, S.~Chakraborty,
  \href{https://doi.org/10.1088/1361-6528/ac6529}{Progress and challenges in
  layered two-dimensional hybrid perovskites}, Nanotechnology 33~(29) (2022)
  292501.
\newblock \href {https://doi.org/10.1088/1361-6528/ac6529}
  {\path{doi:10.1088/1361-6528/ac6529}}.
\newline\urlprefix\url{https://doi.org/10.1088/1361-6528/ac6529}

\bibitem{Ren}
X.~Ren, X.~Yan, D.~V. Gennep, H.~Cheng, L.~Wang, Y.~Li, Y.~Zhao, S.~Wang,
  \href{https://doi.org/10.1063/1.5143795}{Bandgap widening by pressure-induced
  disorder in two-dimensional lead halide perovskite}, Applied Physics Letters
  116~(10) (2020) 101901.
\newblock \href {http://arxiv.org/abs/https://doi.org/10.1063/1.5143795}
  {\path{arXiv:https://doi.org/10.1063/1.5143795}}, \href
  {https://doi.org/10.1063/1.5143795} {\path{doi:10.1063/1.5143795}}.
\newline\urlprefix\url{https://doi.org/10.1063/1.5143795}

\bibitem{Yin}
T.~Yin, B.~Liu, J.~Yan, Y.~Fang, M.~Chen, W.~K. Chong, S.~Jiang, J.-L. Kuo,
  J.~Fang, P.~Liang, S.~Wei, K.~P. Loh, T.~C. Sum, T.~J. White, Z.~X. Shen,
  \href{https://doi.org/10.1021/jacs.8b07765}{Pressure-engineered structural
  and optical properties of two-dimensional (c4h9nh3)2pbi4 perovskite
  exfoliated nm-thin flakes}, Journal of the American Chemical Society 141~(3)
  (2019) 1235--1241.
\newblock \href {http://arxiv.org/abs/https://doi.org/10.1021/jacs.8b07765}
  {\path{arXiv:https://doi.org/10.1021/jacs.8b07765}}, \href
  {https://doi.org/10.1021/jacs.8b07765} {\path{doi:10.1021/jacs.8b07765}}.
\newline\urlprefix\url{https://doi.org/10.1021/jacs.8b07765}

\bibitem{Richard}
G.~Liu, L.~Kong, P.~Guo, C.~C. Stoumpos, Q.~Hu, Z.~Liu, Z.~Cai, D.~J. Gosztola,
  H.-k. Mao, M.~G. Kanatzidis, R.~D. Schaller,
  \href{https://doi.org/10.1021/acsenergylett.7b00807}{Two regimes of bandgap
  red shift and partial ambient retention in pressure-treated two-dimensional
  perovskites}, ACS Energy Letters 2~(11) (2017) 2518--2524.
\newblock \href
  {http://arxiv.org/abs/https://doi.org/10.1021/acsenergylett.7b00807}
  {\path{arXiv:https://doi.org/10.1021/acsenergylett.7b00807}}, \href
  {https://doi.org/10.1021/acsenergylett.7b00807}
  {\path{doi:10.1021/acsenergylett.7b00807}}.
\newline\urlprefix\url{https://doi.org/10.1021/acsenergylett.7b00807}

\bibitem{Quan}
Q.~Li, L.~Yin, Z.~Chen, K.~Deng, S.~Luo, B.~Zou, Z.~Wang, J.~Tang, Z.~Quan,
  \href{https://doi.org/10.1021/acs.inorgchem.8b03190}{High pressure structural
  and optical properties of two-dimensional hybrid halide perovskite
  (ch3nh3)3bi2br9}, Inorganic Chemistry 58~(2) (2019) 1621--1626.
\newblock \href
  {http://arxiv.org/abs/https://doi.org/10.1021/acs.inorgchem.8b03190}
  {\path{arXiv:https://doi.org/10.1021/acs.inorgchem.8b03190}}, \href
  {https://doi.org/10.1021/acs.inorgchem.8b03190}
  {\path{doi:10.1021/acs.inorgchem.8b03190}}.
\newline\urlprefix\url{https://doi.org/10.1021/acs.inorgchem.8b03190}

\bibitem{Mario}
V.~Gómez, S.~Klyatskaya, O.~Fuhr, S.~Kalytchuk, R.~Zbořil, M.~Kappes,
  S.~Lebedkin, M.~Ruben,
  \href{https://doi.org/10.1021/acs.inorgchem.0c01490}{Pressure-modulated
  broadband emission in 2d layered hybrid perovskite-like bromoplumbate},
  Inorganic Chemistry 59~(17) (2020) 12431--12436, pMID: 32838516.
\newblock \href
  {http://arxiv.org/abs/https://doi.org/10.1021/acs.inorgchem.0c01490}
  {\path{arXiv:https://doi.org/10.1021/acs.inorgchem.0c01490}}, \href
  {https://doi.org/10.1021/acs.inorgchem.0c01490}
  {\path{doi:10.1021/acs.inorgchem.0c01490}}.
\newline\urlprefix\url{https://doi.org/10.1021/acs.inorgchem.0c01490}

\bibitem{Marta}
M.~Morana, R.~Chiara, B.~Joseph, T.~B. Shiell, T.~A. Strobel, M.~Coduri,
  G.~Accorsi, A.~Tuissi, A.~Simbula, F.~Pitzalis, A.~Mura, G.~Bongiovanni,
  L.~Malavasi,
  \href{https://www.sciencedirect.com/science/article/pii/S2589004222003273}{Pressure
  response of decylammonium-containing 2d iodide perovskites}, iScience 25~(4)
  (2022) 104057.
\newblock \href {https://doi.org/https://doi.org/10.1016/j.isci.2022.104057}
  {\path{doi:https://doi.org/10.1016/j.isci.2022.104057}}.
\newline\urlprefix\url{https://www.sciencedirect.com/science/article/pii/S2589004222003273}

\bibitem{Dou2015}
L.~Dou, A.~B. Wong, Y.~Yu, M.~Lai, N.~Kornienko, S.~W. Eaton, A.~Fu, C.~G.
  Bischak, J.~Ma, T.~Ding, N.~S. Ginsberg, L.-W. Wang, A.~P. Alivisatos,
  P.~Yang,
  \href{https://www.science.org/doi/abs/10.1126/science.aac7660}{Atomically
  thin two-dimensional organic-inorganic hybrid perovskites}, Science
  349~(6255) (2015) 1518--1521.
\newblock \href
  {http://arxiv.org/abs/https://www.science.org/doi/pdf/10.1126/science.aac7660}
  {\path{arXiv:https://www.science.org/doi/pdf/10.1126/science.aac7660}}, \href
  {https://doi.org/10.1126/science.aac7660}
  {\path{doi:10.1126/science.aac7660}}.
\newline\urlprefix\url{https://www.science.org/doi/abs/10.1126/science.aac7660}

\bibitem{Burschka2013}
J.~Burschka, N.~Pellet, S.-J. Moon, R.~Humphry-Baker, P.~Gao, M.~K.
  Nazeeruddin, M.~Gr{\"a}tzel,
  \href{https://doi.org/10.1038/nature12340}{Sequential deposition as a route
  to high-performance perovskite-sensitized solar cells}, Nature 499~(7458)
  (2013) 316--319.
\newblock \href {https://doi.org/10.1038/nature12340}
  {\path{doi:10.1038/nature12340}}.
\newline\urlprefix\url{https://doi.org/10.1038/nature12340}

\bibitem{Lee2012}
M.~M. Lee, J.~Teuscher, T.~Miyasaka, T.~N. Murakami, H.~J. Snaith,
  \href{https://www.science.org/doi/abs/10.1126/science.1228604}{Efficient
  hybrid solar cells based on meso-superstructured organometal halide
  perovskites}, Science 338~(6107) (2012) 643--647.
\newblock \href
  {http://arxiv.org/abs/https://www.science.org/doi/pdf/10.1126/science.1228604}
  {\path{arXiv:https://www.science.org/doi/pdf/10.1126/science.1228604}}, \href
  {https://doi.org/10.1126/science.1228604}
  {\path{doi:10.1126/science.1228604}}.
\newline\urlprefix\url{https://www.science.org/doi/abs/10.1126/science.1228604}

\bibitem{Nafday2019}
D.~Nafday, D.~Sen, N.~Kaushal, A.~Mukherjee, T.~Saha-Dasgupta,
  \href{https://link.aps.org/doi/10.1103/PhysRevResearch.1.032034}{2d
  ferromagnetism in layered inorganic-organic hybrid perovskites}, Phys. Rev.
  Research 1 (2019) 032034.
\newblock \href {https://doi.org/10.1103/PhysRevResearch.1.032034}
  {\path{doi:10.1103/PhysRevResearch.1.032034}}.
\newline\urlprefix\url{https://link.aps.org/doi/10.1103/PhysRevResearch.1.032034}

\bibitem{Polyakov2012}
A.~O. Polyakov, A.~H. Arkenbout, J.~Baas, G.~R. Blake, A.~Meetsma, A.~Caretta,
  P.~H.~M. van Loosdrecht, T.~T.~M. Palstra,
  \href{https://doi.org/10.1021/cm2023696}{Coexisting ferromagnetic and
  ferroelectric order in a cucl4-based organic--inorganic hybrid}, Chemistry of
  Materials 24~(1) (2012) 133--139.
\newblock \href {https://doi.org/10.1021/cm2023696}
  {\path{doi:10.1021/cm2023696}}.
\newline\urlprefix\url{https://doi.org/10.1021/cm2023696}

\bibitem{Gong2017}
C.~Gong, L.~Li, Z.~Li, H.~Ji, A.~Stern, Y.~Xia, T.~Cao, W.~Bao, C.~Wang,
  Y.~Wang, Z.~Q. Qiu, R.~J. Cava, S.~G. Louie, J.~Xia, X.~Zhang,
  \href{https://doi.org/10.1038/nature22060}{Discovery of intrinsic
  ferromagnetism in two-dimensional van der waals crystals}, Nature 546~(7657)
  (2017) 265--269.
\newblock \href {https://doi.org/10.1038/nature22060}
  {\path{doi:10.1038/nature22060}}.
\newline\urlprefix\url{https://doi.org/10.1038/nature22060}

\bibitem{Mermin66}
N.~D. Mermin, H.~Wagner,
  \href{https://link.aps.org/doi/10.1103/PhysRevLett.17.1133}{Absence of
  ferromagnetism or antiferromagnetism in one- or two-dimensional isotropic
  heisenberg models}, Phys. Rev. Lett. 17 (1966) 1133--1136.
\newblock \href {https://doi.org/10.1103/PhysRevLett.17.1133}
  {\path{doi:10.1103/PhysRevLett.17.1133}}.
\newline\urlprefix\url{https://link.aps.org/doi/10.1103/PhysRevLett.17.1133}

\bibitem{Banerjee_2021_2D}
H.~Banerjee, P.~Barone, S.~Picozzi,
  \href{https://doi.org/10.1088/2053-1583/abe4ba}{Half-metallic ferromagnetism
  in layered {CdOHCl} induced by hole doping}, 2D Materials 8~(2) (2021)
  025027.
\newblock \href {https://doi.org/10.1088/2053-1583/abe4ba}
  {\path{doi:10.1088/2053-1583/abe4ba}}.
\newline\urlprefix\url{https://doi.org/10.1088/2053-1583/abe4ba}

\bibitem{Ohkoshi2011}
S.-i. Ohkoshi, K.~Imoto, Y.~Tsunobuchi, S.~Takano, H.~Tokoro,
  \href{https://doi.org/10.1038/nchem.1067}{Light-induced spin-crossover
  magnet}, Nature Chemistry 3~(7) (2011) 564--569.
\newblock \href {https://doi.org/10.1038/nchem.1067}
  {\path{doi:10.1038/nchem.1067}}.
\newline\urlprefix\url{https://doi.org/10.1038/nchem.1067}

\bibitem{balde}
C.~Balde, W.~Bauer, E.~Kaps, S.~Neville, C.~Desplanches, G.~Chastanet,
  B.~Weber, J.~F. Letard, Light-induced excited spin-state properties in 1d
  iron(ii) chain compounds, European Journal of Inorganic Chemistry 2013~(15)
  (2013) 2744--2750.
\newblock \href {https://doi.org/10.1002/ejic.201201422}
  {\path{doi:10.1002/ejic.201201422}}.

\bibitem{hauser}
A.~Hauser, \href{https://doi.org/10.1063/1.459851}{Intersystem crossing in the
  [fe(ptz)6](bf4)2 spin crossover system (ptz=1‐propyltetrazole)}, The
  Journal of Chemical Physics 94~(4) (1991) 2741--2748.
\newblock \href {http://arxiv.org/abs/https://doi.org/10.1063/1.459851}
  {\path{arXiv:https://doi.org/10.1063/1.459851}}, \href
  {https://doi.org/10.1063/1.459851} {\path{doi:10.1063/1.459851}}.
\newline\urlprefix\url{https://doi.org/10.1063/1.459851}

\bibitem{anna}
G.~D'Avino, A.~Painelli, K.~Boukheddaden,
  \href{https://link.aps.org/doi/10.1103/PhysRevB.84.104119}{Vibronic model for
  spin crossover complexes}, Phys. Rev. B 84 (2011) 104119.
\newblock \href {https://doi.org/10.1103/PhysRevB.84.104119}
  {\path{doi:10.1103/PhysRevB.84.104119}}.
\newline\urlprefix\url{https://link.aps.org/doi/10.1103/PhysRevB.84.104119}

\bibitem{ramasesha}
R.~Raghunathan, S.~Ramasesha, C.~Mathoniere, V.~Marvaud, Microscopic model for
  photoinduced magnetism in the molecular complex
  [mo(iv)(cn)$_2$(cn-cul)$_6$]$^{8+}$ perchlorate, Phys. Rev. B 73 (2006)
  045131.
\newblock \href {https://doi.org/10.1103/PhysRevB.73.045131}
  {\path{doi:10.1103/PhysRevB.73.045131}}.

\bibitem{daubric}
H.~Daubric, R.~Berger, J.~Kliava, G.~Chastanet, O.~Nguyen, J.-F. L\'etard,
  \href{https://link.aps.org/doi/10.1103/PhysRevB.66.054423}{Light-induced
  excited spin-state trapping of ${\mathrm{fe}}^{2+}$ observed by electron
  paramagnetic resonance of ${\mathrm{mn}}^{2+}$}, Phys. Rev. B 66 (2002)
  054423.
\newblock \href {https://doi.org/10.1103/PhysRevB.66.054423}
  {\path{doi:10.1103/PhysRevB.66.054423}}.
\newline\urlprefix\url{https://link.aps.org/doi/10.1103/PhysRevB.66.054423}

\bibitem{Marshak2012}
M.~P. Marshak, M.~B. Chambers, D.~G. Nocera,
  \href{https://doi.org/10.1021/ic301970w}{Cobalt in a bis-$\beta$-diketiminate
  environment}, Inorganic Chemistry 51~(20) (2012) 11190--11197.
\newblock \href {https://doi.org/10.1021/ic301970w}
  {\path{doi:10.1021/ic301970w}}.
\newline\urlprefix\url{https://doi.org/10.1021/ic301970w}

\bibitem{hauser2}
A.~Hauser, A.~Vef, P.~Adler,
  \href{https://doi.org/10.1063/1.461255}{Intersystem crossing dynamics in
  fe(ii) coordination compounds}, The Journal of Chemical Physics 95~(12)
  (1991) 8710--8717.
\newblock \href {http://arxiv.org/abs/https://doi.org/10.1063/1.461255}
  {\path{arXiv:https://doi.org/10.1063/1.461255}}, \href
  {https://doi.org/10.1063/1.461255} {\path{doi:10.1063/1.461255}}.
\newline\urlprefix\url{https://doi.org/10.1063/1.461255}

\bibitem{Liu2013}
T.~Liu, H.~Zheng, S.~Kang, Y.~Shiota, S.~Hayami, M.~Mito, O.~Sato,
  K.~Yoshizawa, S.~Kanegawa, C.~Duan, A light-induced spin crossover actuated
  single-chain magnet, Nature Communications 4~(1) (2013) 2826.
\newblock \href {https://doi.org/10.1038/ncomms3826}
  {\path{doi:10.1038/ncomms3826}}.

\bibitem{Bertoni2015}
R.~Bertoni, M.~Cammarata, M.~Lorenc, S.~F. Matar, J.-F. L{\'e}tard, H.~T.
  Lemke, E.~Collet, \href{https://doi.org/10.1021/ar500444d}{Ultrafast
  light-induced spin-state trapping photophysics investigated in
  fe(phen)2(ncs)2 spin-crossover crystal}, Accounts of Chemical Research 48~(3)
  (2015) 774--781.
\newblock \href {https://doi.org/10.1021/ar500444d}
  {\path{doi:10.1021/ar500444d}}.
\newline\urlprefix\url{https://doi.org/10.1021/ar500444d}

\bibitem{GUTLICH19901}
P.~Gutlich, A.~Hauser,
  \href{http://www.sciencedirect.com/science/article/pii/0010854590800766}{Thermal
  and light-induced spin crossover in iron(ii) complexes}, Coordination
  Chemistry Reviews 97 (1990) 1 -- 22.
\newblock \href {https://doi.org/https://doi.org/10.1016/0010-8545(90)80076-6}
  {\path{doi:https://doi.org/10.1016/0010-8545(90)80076-6}}.
\newline\urlprefix\url{http://www.sciencedirect.com/science/article/pii/0010854590800766}

\bibitem{Karmakar2022}
S.~Karmakar, P.~Chakraborty, T.~Saha-Dasgupta,
  \href{http://dx.doi.org/10.1039/D2CP00539E}{Trend in light-induced
  excited-state spin trapping in fe(ii)-based spin crossover systems}, Phys.
  Chem. Chem. Phys. 24 (2022) 10201--10209.
\newblock \href {https://doi.org/10.1039/D2CP00539E}
  {\path{doi:10.1039/D2CP00539E}}.
\newline\urlprefix\url{http://dx.doi.org/10.1039/D2CP00539E}

\bibitem{Unger}
E.~L. Unger, E.~T. Hoke, C.~D. Bailie, W.~H. Nguyen, A.~R. Bowring,
  T.~Heumüller, M.~G. Christoforo, M.~D. McGehee,
  \href{http://dx.doi.org/10.1039/C4EE02465F}{Hysteresis and transient behavior
  in current–voltage measurements of hybrid-perovskite absorber solar cells},
  Energy Environ. Sci. 7 (2014) 3690--3698.
\newblock \href {https://doi.org/10.1039/C4EE02465F}
  {\path{doi:10.1039/C4EE02465F}}.
\newline\urlprefix\url{http://dx.doi.org/10.1039/C4EE02465F}

\bibitem{Barrows}
A.~T. Barrows, A.~J. Pearson, C.~K. Kwak, A.~D.~F. Dunbar, A.~R. Buckley, D.~G.
  Lidzey, \href{http://dx.doi.org/10.1039/C4EE01546K}{Efficient planar
  heterojunction mixed-halide perovskite solar cells deposited via
  spray-deposition}, Energy Environ. Sci. 7 (2014) 2944--2950.
\newblock \href {https://doi.org/10.1039/C4EE01546K}
  {\path{doi:10.1039/C4EE01546K}}.
\newline\urlprefix\url{http://dx.doi.org/10.1039/C4EE01546K}

\bibitem{Liu}
C.~Liu, J.~Fan, X.~Zhang, Y.~Shen, L.~Yang, Y.~Mai,
  \href{https://doi.org/10.1021/acsami.5b00375}{Hysteretic behavior upon light
  soaking in perovskite solar cells prepared via modified vapor-assisted
  solution process}, ACS Applied Materials \& Interfaces 7~(17) (2015)
  9066--9071, pMID: 25860158.
\newblock \href {http://arxiv.org/abs/https://doi.org/10.1021/acsami.5b00375}
  {\path{arXiv:https://doi.org/10.1021/acsami.5b00375}}, \href
  {https://doi.org/10.1021/acsami.5b00375} {\path{doi:10.1021/acsami.5b00375}}.
\newline\urlprefix\url{https://doi.org/10.1021/acsami.5b00375}

\bibitem{Zhao}
C.~Zhao, B.~Chen, X.~Qiao, L.~Luan, K.~Lu, B.~Hu,
  \href{https://onlinelibrary.wiley.com/doi/abs/10.1002/aenm.201500279}{Revealing
  underlying processes involved in light soaking effects and hysteresis
  phenomena in perovskite solar cells}, Advanced Energy Materials 5~(14) (2015)
  1500279.
\newblock \href
  {http://arxiv.org/abs/https://onlinelibrary.wiley.com/doi/pdf/10.1002/aenm.201500279}
  {\path{arXiv:https://onlinelibrary.wiley.com/doi/pdf/10.1002/aenm.201500279}},
  \href {https://doi.org/https://doi.org/10.1002/aenm.201500279}
  {\path{doi:https://doi.org/10.1002/aenm.201500279}}.
\newline\urlprefix\url{https://onlinelibrary.wiley.com/doi/abs/10.1002/aenm.201500279}

\bibitem{Kim2019}
D.~Kim, J.~S. Yun, P.~Sharma, D.~S. Lee, J.~Kim, A.~M. Soufiani, S.~Huang,
  M.~A. Green, A.~W.~Y. Ho-Baillie, J.~Seidel,
  \href{https://doi.org/10.1038/s41467-019-08364-1}{Light- and bias-induced
  structural variations in metal halide perovskites}, Nature Communications
  10~(1) (2019) 444.
\newblock \href {https://doi.org/10.1038/s41467-019-08364-1}
  {\path{doi:10.1038/s41467-019-08364-1}}.
\newline\urlprefix\url{https://doi.org/10.1038/s41467-019-08364-1}

\bibitem{Kepenekian2015}
M.~Kepenekian, R.~Robles, C.~Katan, D.~Sapori, L.~Pedesseau, J.~Even,
  \href{https://doi.org/10.1021/acsnano.5b04409}{Rashba and dresselhaus effects
  in hybrid organic-inorganic perovskites: From basics to devices}, ACS Nano
  9~(12) (2015) 11557--11567.
\newblock \href {https://doi.org/10.1021/acsnano.5b04409}
  {\path{doi:10.1021/acsnano.5b04409}}.
\newline\urlprefix\url{https://doi.org/10.1021/acsnano.5b04409}

\bibitem{Kepenekian2017}
M.~Kepenekian, J.~Even,
  \href{https://doi.org/10.1021/acs.jpclett.7b01015}{Rashba and dresselhaus
  couplings in halide perovskites: Accomplishments and opportunities for
  spintronics and spin-orbitronics}, The Journal of Physical Chemistry Letters
  8~(14) (2017) 3362--3370.
\newblock \href {https://doi.org/10.1021/acs.jpclett.7b01015}
  {\path{doi:10.1021/acs.jpclett.7b01015}}.
\newline\urlprefix\url{https://doi.org/10.1021/acs.jpclett.7b01015}

\end{thebibliography}

\end{document}